%% file: vldb2019.tex
\documentclass{vldb}
\usepackage{graphicx}
\usepackage{balance}  
\usepackage{subcaption}
\usepackage{xcolor}
\usepackage{xspace}
\usepackage{cite}
\usepackage{booktabs} 

\newcommand{\sparagraph}[1]{\noindent {\bf #1}}

\newcommand{\neural}{{Neural}\xspace}
\newcommand{\pca}{{PCA}\xspace}
\newcommand{\fa}{{FA}\xspace}
\newcommand{\sparse}{{Sparse}\xspace}

\definecolor{lblue}{HTML}{8888FF}

\definecolor{darkblue}{HTML}{0000A0}
\definecolor{dyellow}{HTML}{CCCC00}

\definecolor{darkgreen}{HTML}{006400}
\definecolor{do}{HTML}{9932CC}

\vldbTitle{Flexible Operator Embeddings via Deep Learning}
\vldbAuthors{Ryan Marcus, Olga Papaemmanouil}
\vldbDOI{https://doi.org/10.14778/xxxxxxx.xxxxxxx}
\vldbVolume{12}
\vldbNumber{xxx}
\vldbYear{2019}

\begin{document}


\title{Flexible Operator Embeddings via Deep Learning}



%
%
%
%

\numberofauthors{2} 

\author{
%
%
\alignauthor
Ryan Marcus\\
       \affaddr{Brandeis University}\\
       \email{ryan@cs.brandeis.edu}
\alignauthor
 Olga Papaemmanouil\\
       \affaddr{Brandeis University}\\
       \email{olga@cs.brandeis.edu}
}


\toappear{}
\maketitle

\begin{abstract}
Integrating machine learning techniques into the internals of database systems requires significant feature engineering, a human effort-intensive process to determine the best way to represent relevant pieces of information for a data management task. In addition to being labor intensive, the process of hand-engineering features must generally be repeated for each task, and may make assumptions about the underlying database that are not universally true. We introduce a deep learning technique for \emph{automatically} transforming query operators into dense, information-rich feature vectors, called operator embeddings. These learned embeddings are custom-tailored to the underlying database, enabling off-the-shelf machine learning techniques to achieve good performance on multiple data management tasks. Experimentally, we show that our flexible operator embeddings outperform existing feature engineering techniques using both synthetic and real-world datasets.
\end{abstract}

\input{introduction}

\input{embeddings}

\input{system}
\input{training}
\input{experiments}
\input{related_work}
\input{conclusion}



\appendix
\input{apx_encoding}

\clearpage
\bibliographystyle{abbrv}
\bibliography{ryan-cites-short}  

\end{document}

%% file: introduction.tex
\section{Introduction}

As database management systems and their applications grow increasingly complex, researchers have turned to machine learning to solve numerous data management problems, such as query admission control~\cite{q-cop}, query scheduling~\cite{wisedb-vldb}, cluster sizing~\cite{perfenforce_demo, mdp_elastic, wisedb-cidr}, index selection~\cite{selfdrivingcidr}, cardinality estimation~\cite{deep_card_est, deep_card_est2}, and query optimization~\cite{rejoin, sanjay_wat, qo_state_rep}. One of the most critical steps of integrating machine learning techniques into data management systems is the process of \emph{feature engineering}.  Generally, feature engineering involves human experts deciding (1) if a particular piece of information is relevant to a task (too many features may increase inference time or decrease the efficacy of some machine learning models), (2) how to \emph{combine} available domain information into \emph{useful} features (many models desirable for their fast inference time are unable to learn arbitrary combinations of features, e.g. linear regression), and (3) how to \emph{encode} these features into \emph{usable} input (most machine learning algorithms require vector inputs). 

In addition to being a difficult task to undertake, feature engineering has many disadvantages specific to data management tasks:

\begin{itemize}
\item{{\bf Time/cost intensive}: feature engineering requires experts in machine learning and database systems to spend a long time testing and experimenting with different combinations of features. For example, there were 18 engineered features in~\cite{learning_latency} for latency prediction, and  41 engineered features in~\cite{Khannewintrusiondetection2007} used for database intrusion detection.}
\item{{\bf Non-task transferable}: generally, features engineered by hand for a specific task will not provide good performance on a different task -- the process must be repeated for each task (e.g. the features developed for query admission control in~\cite{q-cop} cannot be used for latency prediction~\cite{learning_latency}, nor for database intrusion detection~\cite{Khannewintrusiondetection2007}, and vice versa).}
\item{{\bf Non-data transferable}: features engineered for a particular task may work well for one dataset (e.g. TPC-H), but those same features may fail for another dataset (e.g. a real-world dashboard system).}
\end{itemize}

In order to ease these burdens, this work introduces \emph{flexible operator embeddings}: a general technique to perform multi-purpose, database-tailored feature engineering with minimal human effort. As query operators lie at the heart of numerous complex tasks, including but not limited to resource consumption prediction~\cite{LiRobustestimationresource2012}, query optimization~\cite{systemr}, query performance prediction~\cite{learning_latency}, and query scheduling~\cite{q-cop}, our embeddings offer feature engineering at the operator level. Here, an \emph{operator embedding} is a mapping from a query operator to a low-dimensional feature (vector) space that captures rich, useful information about the operator. By leveraging deep learning techniques, these embeddings act as vectorized representations of an operator that (a) can be generated and tailored to a specific database, (b)  can provide useful features for a variety of data management tasks, and (c) provide usable input to numerous off-the-shelf machine learning algorithms.

Our technique for learning representations of query operators is based on the intuition that the behavior of an operator is \emph{context sensitive}: a join of two large relations will behave differently than a join of two small relations (e.g. greater memory usage, possible buffer spilling, etc.). We capture this intuition by training a deep neural network with a specialized, hour-glass architecture to predict the context of a given query operator: in our case, the operator's children.  The training set for this neural network is a set of query plans (i.e., trees of query operators). Once trained, we use the internal representation of a particular operator learned by the neural network as that operator's embedding. Intuitively, training the neural network to predict an operator's children ensures that the internal representation learned by the neural network carries a high amount of semantic information about the operator. For example, if the network can predict that a join operator with a particular predicate normally has two very large inputs, then the internal representation learned by the network is likely to carry semantic information relevant to whether or not that join operator will use a significant amount of memory.

One unique characteristic of our operator embeddings is that they produce features that can be fed to traditional, off-the-shelf machine learning models with fast inference time. Furthermore, they can be useful for integrating machine learning techniques into  a variety of data management tasks such as  query admission, query outlier detection, and detecting cardinality estimation errors. Finally, once learned by the neural network, our flexible operator embeddings are significantly information rich. This implies that, when learned embeddings are used as input to a traditional machine learning model, only a relatively small amount of training data is required to learn a particular task. Overall, we argue that learned embeddings represent a valuable addition to the DBMS designer's toolbox, enabling accurate, light-weight, custom-tailored models for a wide variety of tasks that can be generated with minimal human effort.

The contributions of this paper are:
\begin{itemize}
\item{We present \emph{flexible operator embeddings}, a novel technique to automatically map query operators to useful features tailored to a particular database, alleviating the need for costly human feature engineering.}
\item{We demonstrate that our embeddings can be effectively combined with simple, low-latency, and off-the-shelf machine learning models.}
  \item{We demonstrate the efficacy and flexibility of our custom embeddings across a variety of datasets and tasks.}
\end{itemize}

This remainder of this paper is organized as follows. In Section~\ref{sec:embedding}, we give an intuitive overview of why effectively representing query operators as vectors is difficult, and how deep learning can be used to learn effective embeddings. Section~\ref{sec:system} describes our framework for learning and applying database-tailored operator embeddings. Section~\ref{sec:training} describes how to train a custom operator embedding and apply it to a particular task. We present experimental results in Section~\ref{sec:experiments}, demonstrating the efficacy of our approach across a variety of workloads and tasks. Related work is presented in Section~\ref{sec:related}, and concluding remarks are given in Section~\ref{sec:conclusions}.


%% file: embeddings.tex
\section{Context-aware Embeddings}
\label{sec:embedding}

Machine learning is not magic, and like all algorithms, obeys the maxim of ``garbage in, garbage out.'' If one provides a meaningless or nearly-meaningless representation of the domain, a machine learning algorithm will show poor performance~\cite{gigo}. If one wants a machine learning model to generalize well to unseen data, one needs to provide a good representation of the data. Here, ``good'' inputs to a machine learning algorithm means inputs that are generally \emph{dense} (i.e., every dimension is more-or-less continuous and meaningful) and \emph{information-rich} (i.e., the distance between points is meaningful). Unfortunately, coming up with a dense, information-rich representation of query operators is often not easy. Next, we motivate our context-aware operator embeddings by discussing  alternative approaches and their drawbacks when used for  data management tasks. 

\begin{figure}
  \centering
  \begin{subfigure}{0.4\textwidth}
    \includegraphics[width=\textwidth]{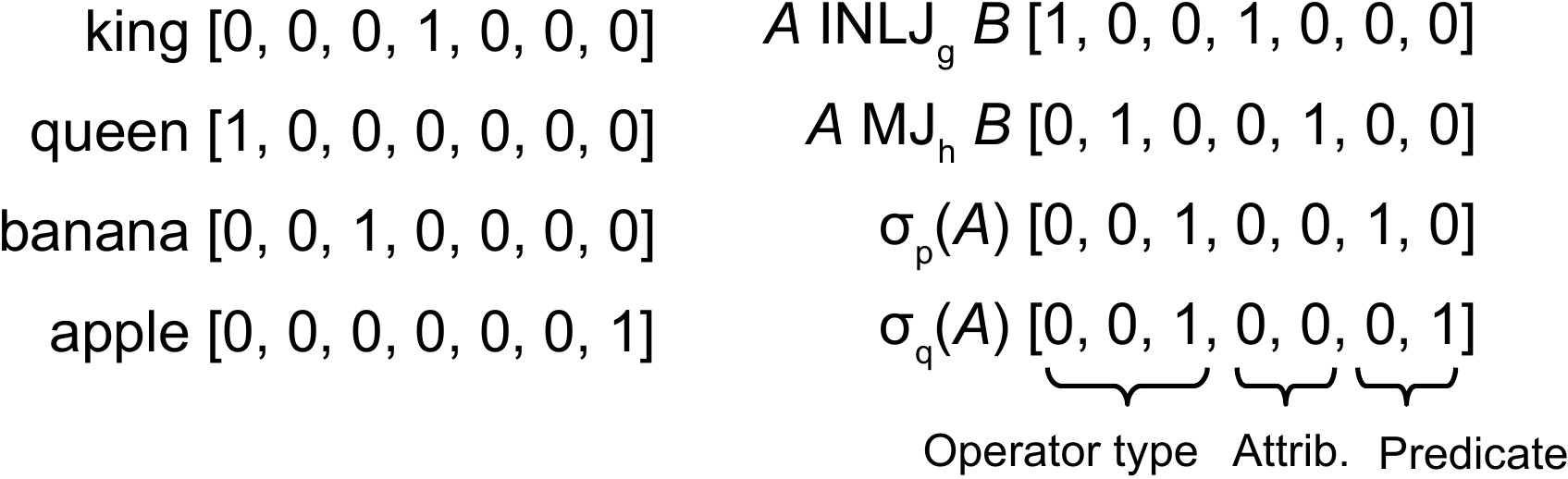}
    \caption{One-hot encodings of words (left) and query operators (right)}
    \vspace{1em}
    \label{fig:w_v_o1}
  \end{subfigure}
  \begin{subfigure}{0.4\textwidth}
    \includegraphics[width=\textwidth]{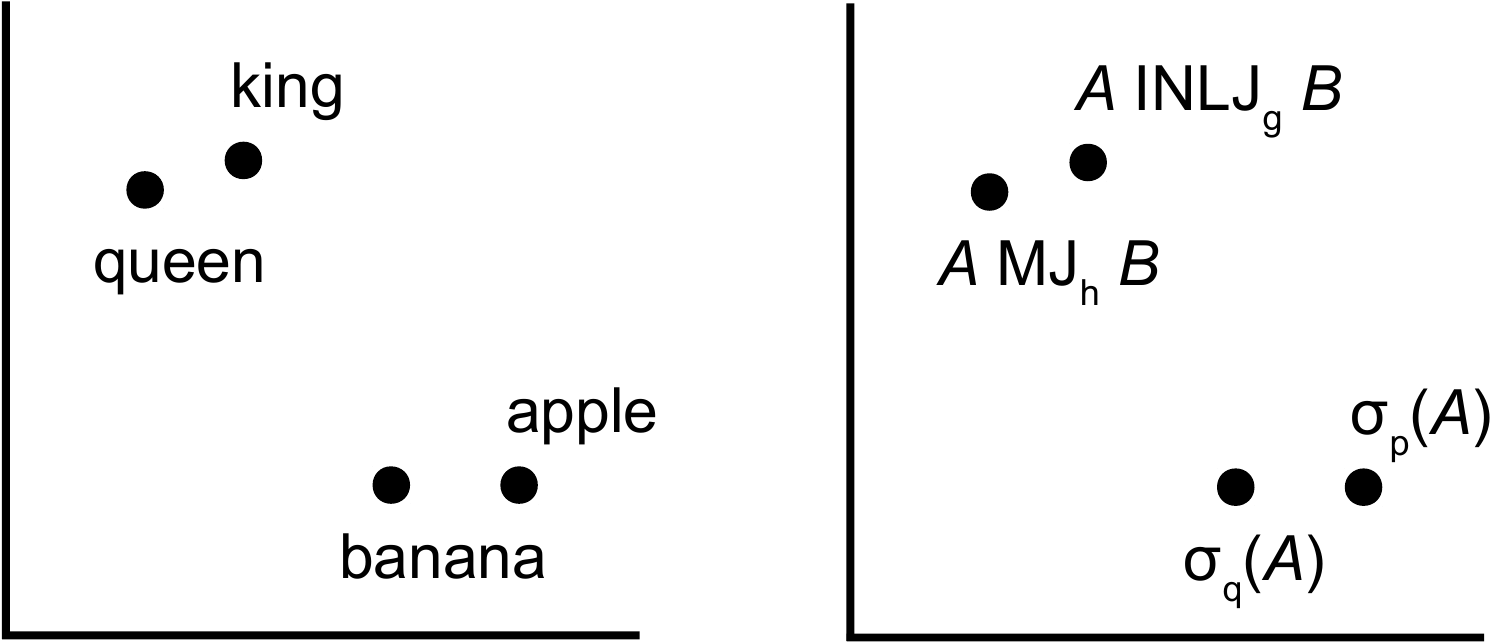}
    \caption{Example embedded space word vectors (left) and operator embeddings (right)}
    \label{fig:w_v_o2}
  \end{subfigure}
  \caption{Encodings and embeddings for words and query operators}
\end{figure}

\sparagraph{One-hot encodings}  A commonly used information encoding approach (particularly in the domain of natural language processing, NLP) is using an one-hot vector, i.e., a vector where there is a single high (generally 1) value and all the other values are low values (generally 0). For example, in NLP, where the main units of analysis are words, a given word can be represented with a high value in a specific position in a vector with the size of the vocabulary.  Figure~\ref{fig:w_v_o1} shows an example of how the words king, queen, apple, and banana might be represented as part of an eight-word vocabulary. Here, the word ``king'' is represented by a high value in the fourth position of the vector, and the word ``queen'' is represented by a high value in the first position.

The right side of Figure~\ref{fig:w_v_o1} shows how a similar one-hot encoding strategy can be used to encode query operators. Here, four operators -- an index-nested loop (natural) join on an attribute $g$, a merge (natural) join on attribute $h$, and two selection operators, with predicates $p$ and $q$ -- are represented with a straightforward ``combined'' one-hot encoding which captures information about the operator type (in the first three dimensions), join attribute (in the next two dimensions), and predicate (in the last two dimensions). For example, the appearance of nested loop join is captured by the value of 1 in the first position, a merge-join is represented by the value of 1 in the second position, and the selection operator is captured in the third position. Similarly, the join attributes ``g'' and``h'' are captured by a value of 1 in the fourth and fifth positions, respectively.


{For both English words and query operators, it is immediately clear that this representation is (1) not dense, as each dimension is used minimally, and (2) not information-rich, as all three operator types (and all four English words) are equidistant  from each other in the embedded space. For example, the distance between the first three operators (the nested loop join, the merge join, and the first selection operator) are all equidistant. As a result, this representation encodes very little semantic information. While not particularly useful (indeed, both database-related and NLP models trained directly on one-hot vectors perform poorly), this sort of encoding is very easy to come up with, and requires almost no human effort.}

{\sparagraph{Information-rich embeddings} Ideally, our vectorized representations would contain significant semantic information. In the NLP case, what is desired is the representation depicted in the left side of Figure~\ref{fig:w_v_o2}: ``king'' and ``queen'' are neighbors, as are ``apple'' and ``banana.'' For query operators, we want the vectors representing the  merge join and index nested loop join  operators to be closer to each other than to the selections, as depicted in the right side of Figure~\ref{fig:w_v_o2}, since both merge join and index nested loop join implement the same logical operator (a join). However, unlike one-hot encodings, it is not trivial to construct such an embedding.} Thus, we ask a natural question: \emph{can we automatically transform the one-hot, sparse, easily-engineered, information-anemic representation into a dense, information-rich representation useful for machine learning algorithms?}

\sparagraph{Context-aware representations} In order to facilitate learning a dense, information-rich representation from an easy-to-construct one-hot encoding, the notion of \emph{context} is often leveraged~\cite{word2vec, wordvec1, wordvec2, wordvec3}.  For instance, \emph{word vectors}, invented to transform words into vectors, is a way to take advantage of the {context} that a word appears in to represent that word as a vector. For example, in the sentence ``Long~live~the~\underline{\ \ \ \ \ \ \ \ }!'', we except to see a word like ``king'' or ``queen'', as opposed to a word like ``apples'' or ``bananas.'' However, in the sentence ``\underline{\ \ \ \ \ \ \ \ }~are~a~tasty~fruit.'', the converse is true: one would expect ``apples'' or ``bananas'' instead of ``king'' or ``queen.''


In a query processing context, if we know that the child of an unknown unary operator is a scan operator reading phone number data, we know it is unlikely that the unknown operator is an average aggregate (as the numerical average of a set of phone numbers is nonsensical). If we know that both the children of an unknown binary operator are sort operators, we know the operator is more likely to be a merge join than a hash join (a query optimizer would have little reason to sort the input to a hash join operator). Hence, in the next paragraphs,  we demonstrate how we can build these context-aware embeddings via deep neural network (DNNs).

\subsection{Context-aware Learning via  DNNs}
Context and deep neural neworks (DNNs) can be leveraged to \emph{learn} a semantically-rich, low-dimensional vector space from sparse, one-hot vectors~\cite{word2vec}.  At a high level, our approach works by training a neural network to predict the context (children) of an operator. In practice, this neural network maps one-hot representations of query operators to one-hot representations of their children. Once trained, the internal representation learned by the neural network is used as an operator embedding.

To facilitate the discussion in the next paragraphs we first provide necessary background on neural networks.  

\sparagraph{Neural networks} Deep neural networks are composed of multiple \emph{layers} of nodes, in which each node takes input from the previous layer, transforms it via a differentiable function and a set of \emph{weights}, and passes the transformed data to the next layer. Thus, the output of each node can be thought of as a \emph{feature}: a piece of information derived from the input data. As you advance into the subsequent layers of the neural net, nodes begin to represent more complex features, since each node aggregates and recombines features from the previous layer. Because each node applies a differentiable transformation, and thus the entire network can be viewed as applying a single complex differentiable transformation, neural networks can be trained using gradient descent~\cite{sgd}. This training is done by defining a \emph{loss function}, a function that measures the prediction accuracy of the neural network on the desired task, and then using gradient descent to modify the parameters (i.e., weights) of the neural network to minimize this loss function. Unlike most traditional machine-learning algorithms, deep neural networks learn complex features \emph{automatically}, without human intervention. Readers unfamiliar with deep neural networks may wish to see~\cite{dnn} for an overview.

\sparagraph{Hour-glass DNN architecture} Next, we discuss how deep neural networks can be used to learn context-aware operator embeddings. 
Our framework takes advantage of contextual relationships between query operators. The critical component is a deep neural network architecture that is trained to predict the context of a particular entity. For example, in the domain of query processing, the neural network, given a merge join operator, would be trained to predict the two sorted inputs (details  in Section~\ref{sec:training}).

Intuitively, this process works by first training a neural network to map a simple, one-hot encoding of a query operator to some contextual information about that operator: for example, the network would be trained to  map a one-hot representation of merge join operator to the one-hot representation of the merge join's two (sorted) inputs. Here, there is no separate process needed apart from training the neural network for this auxiliary task. Once the network is trained,  each hidden layer outputs combinations of its input features that are useful for this predictive task. Hence, these outputs  can be seen an embedding of the (sparse) input vectors of the neural network. 



\begin{figure}
  \centering
  \includegraphics[width=0.48\textwidth]{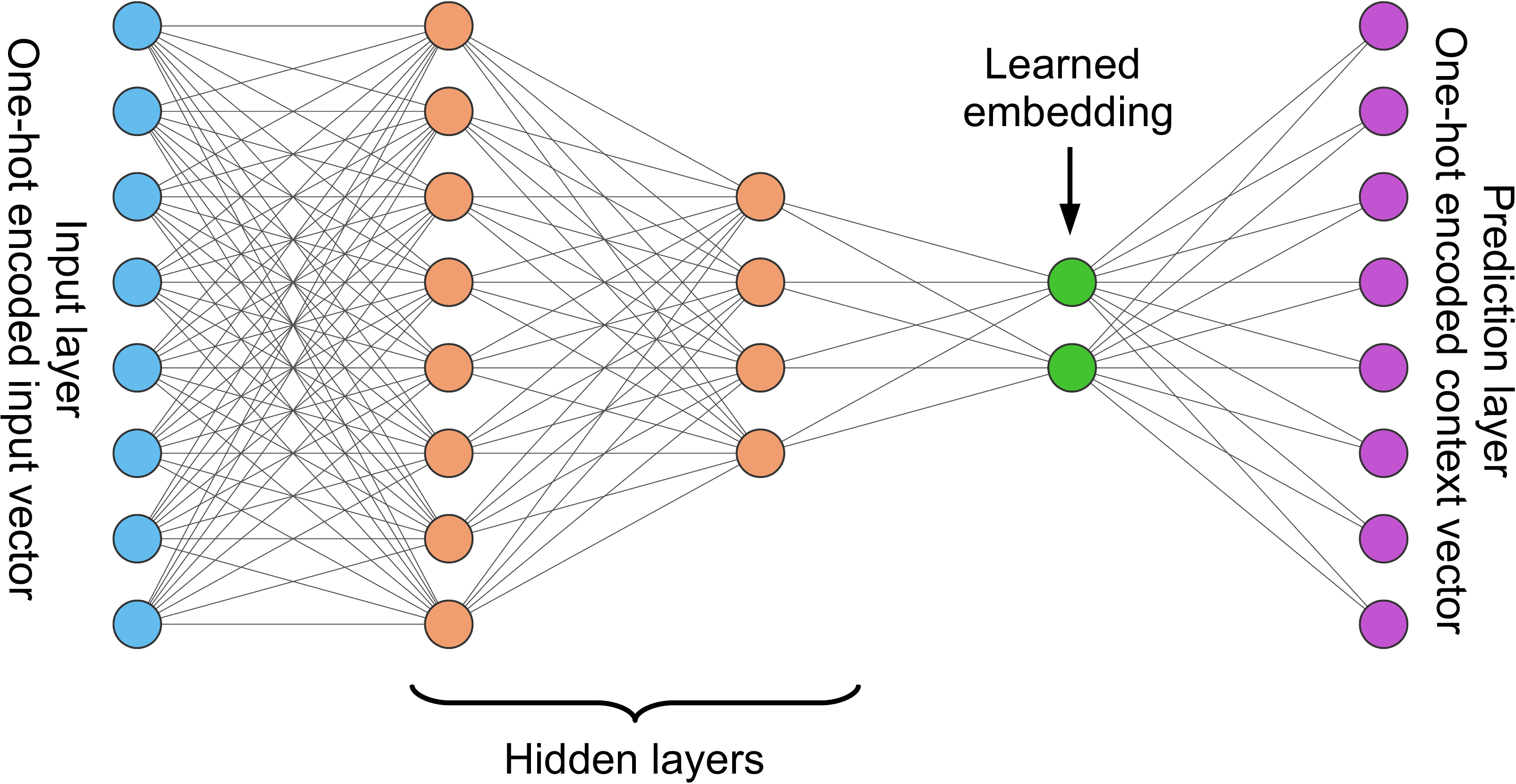}
  \caption{Hourglass embedding neural network} 
  \label{fig:encoder_net}
\end{figure}

To ensure that the final hidden layer of the network offers an output (an embedding) that is dense and information-rich, the architecture of this neural network  is designed with an \emph{information bottleneck}. We use an ``hour-glass'' shaped network, depicted in Figure~\ref{fig:encoder_net}. Here, one-hot vectors of operator information are fed into an \textcolor{darkblue}{input layer} and then projected down into lower and lower dimensional spaces through a series of \textcolor{orange}{hidden layers}. Then, a very low dimensional layer is used to represent the \textcolor{darkgreen}{learned embedding}. At the end of the network, a final \textcolor{purple}{prediction layer} is responsible for mapping the low-dimensional learned embedding into a prediction of the context of the input operator's feature vector.

During training, the neural network strives to identify complex transformations of its input that improve the accuracy of its output prediction. Implicitly, this forces the neural network  to learn a representation (an embedding) of the input one-hot vectors that can be used to predict the correct operator context at the final layer.  After the network is trained, the prediction layer is ``cut off'', resulting in a network which takes in an one-hot encoded, sparse vector and outputs a lower-dimensional embedding. Because this low-dimension learned embedding was trained to be useful for predicting the context of the operator, we know it is information-rich. The low-dimension layer representing the learned embedding serves as an information bottleneck, forcing the learned representation to be dense.


Of course, \emph{precisely} predicting the context of a query operator is an impossible task. After all,  the children of a merge join operator will not always be sort operators. Since ultimately we will cut off the final prediction layer, we do not care much about its accuracy, as long as its output somewhat matches the \emph{distribution} of potential contexts. For example, when fed a merge join operator, the network predicts a higher likelihood that the children of the merge join operator were sort operators than hash operators. Once trained to have this property, the resulting network (with the prediction layer cut off), serves as a mapping from a sparse, information-anemic, easy-to-engineer representation to a dense, information-rich representation. 

The rest of this paper explains how we employ this learned embedding framework to automate and custom-tailor the feature engineering process to a particular database, and how the output of our feature engineering process can be useful for a number of data management tasks.


%% file: system.tex
\section{Learning Framework Overview}
\label{sec:system}
\begin{figure}
  \centering
  \includegraphics[width=0.35\textwidth]{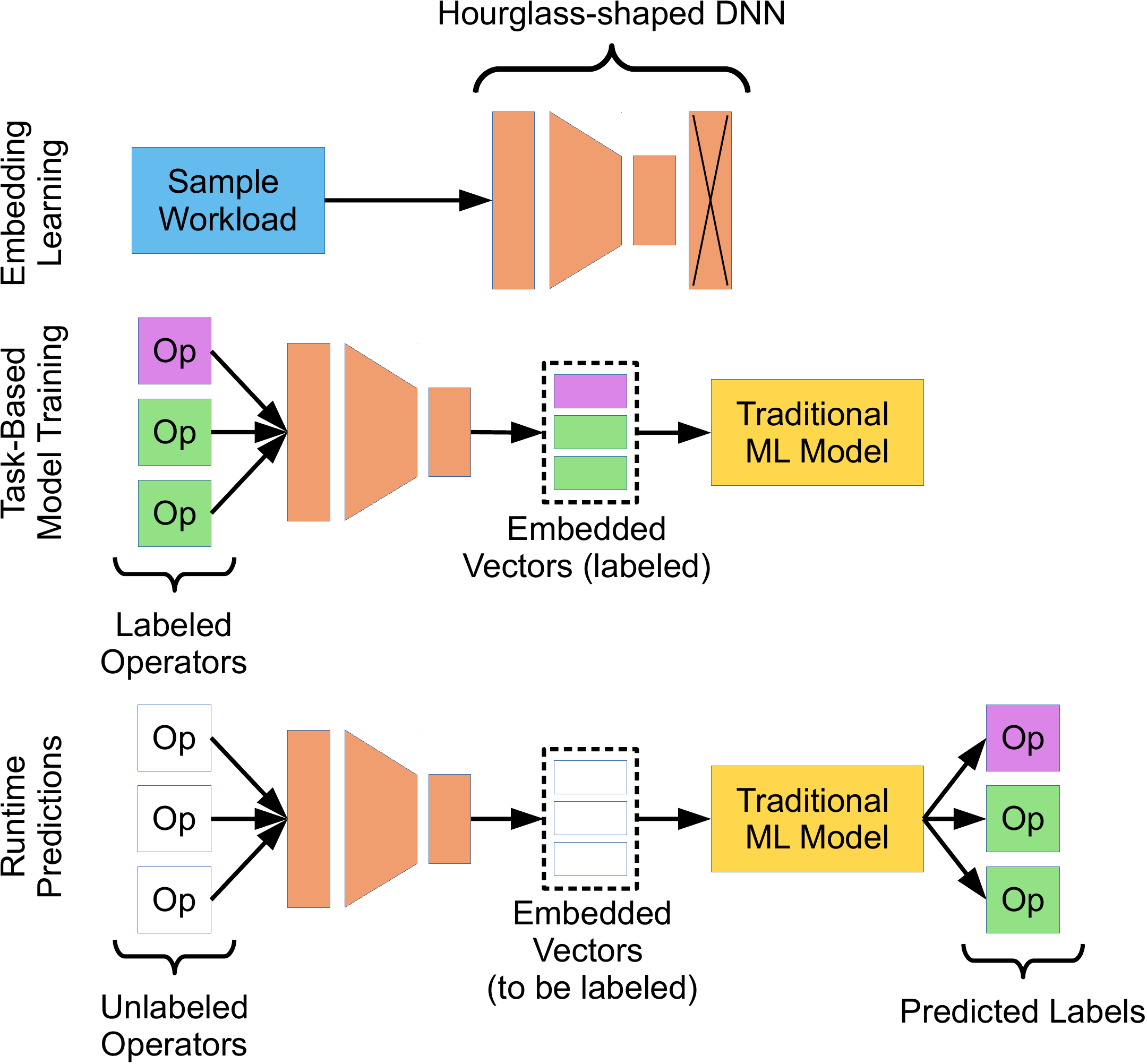}
  \caption{Three phase learning framework}
  \label{fig:system}
\end{figure}

Learning  database-specific operator embeddings is the first step towards automating the feature engineering process. Once operator embeddings are obtained, an off-the-shelf machine learning algorithm can utilize them as an input for a specific task. Next, we provide a brief overview of our framework before describing its technical details in Section~\ref{sec:training}.
 
Our proposed framework operates in three phases, depicted in Figure~\ref{fig:system}.  In the first \emph{embedding learning} step,  an operator embedding is learned by training the ``hour-glass'' neural network. This training uses previously-observed query plans (i.e., operator trees) from a particular database, allowing the network to learn to predict an operator's children (i.e., context). In the  \emph{task-based model training} phase, the learned embeddings are used to train an off-the-shelf machine learning model for a specific task  (i.e., cardinality error estimation). Specifically, the trained cut-off neural network is used to map the training set for this task into the dense embedded vector space and these dense vectors are fed into a traditional machine learning model to train the model for the particular task. Finally, at runtime, observed query operators are embedded through the cut-off neural network and the embedded output is sent through the traditional machine learning model, resulting in a prediction for the particular task and for the observed operator.

\sparagraph{Embedding learning} This first phase focuses on learning the operator embedding for a particular database workload. Here, we assume the availability of a long history of executed queries from the database, which we call a sample workload. This sample workload can be extracted from database logs or through other means. Many DBMSs automatically store such logs for auditing and debugging purposes. The only information needed from these queries are their query plans, which allows one to extract information about each plan's query operators, e.g., the type of operator, the expected cost of the operator according to the optimizer's cost model, etc.\footnote{This information is made available in many DBMSes via \texttt{EXPLAIN} queries.} However, these  queries need no additional annotation or tagging, i.e. they do not have to be hand-labeled with any particular piece of information. 

In this  phase, we train a deep neural network with an hour-glass architecture (orange in Figure~\ref{fig:system}). The first layer  takes as input any available piece of information about a query operator in the training set. The subsequent, hidden layers of the network slowly decrease in size until the penultimate layer of the network, called the \emph{embedding layer}. After the embedding layer, an additional prediction layer is added, responsible for taking the information from the embedding layer and predicting the input operator's children. The embedding layer, the smallest layer of the network, serves as an information bottleneck, forcing the neural network to learn a compact representation of the input operators.

Given the trained neural network, the last step of the embedding learning stage is to ``cut off'' the final prediction layer of the network. This results in a network that maps an input operator to a low-dimensional vector, i.e. the embedding layer becomes the output layer of the network. 


\sparagraph{Task-based model training}
During the model training phase, depicted in the second row of Figure~\ref{fig:system}, the learned embeddings are used to train an off-the-shelf machine learning model from a small training set. For example, suppose one is interested in training a machine learning model to predict cardinality estimation errors, as are common in queries with many joins~\cite{howgood}. One would then collect a small training set consisting of query operators and labels specifying whether or not the estimated cardinality of the operator was too high or too low (these labels are represented with green and purple in Figure~\ref{fig:system}). These collected operators are then encoded into one-hot sparse vectors. These sparse vectors are ran through the hour-glass neural network, generating one dense vector for each operator. These generated vectors, which can be thought of as embedded versions of the labeled operators, are dense and low-dimensional. These vectors, and their corresponding labels, are then used to train an off-the-shelf machine learning model  (e.g.,  logistic regression~\cite{logistic}, random forest~\cite{rf}) to predict whether the candinality estimation of a particular operator will be too high or too low.



\sparagraph{Runtime Predictions}
At runtime, the specialized model trained in the previous step is used on new, unlabeled operators. First, these previously-unseen operators are sent through the cut-off neural network. This produces an embedded dense vector which can be classified by the traditional machine learning model trained in the previous stage. For example, the specialized model may classify the new operator as having an under or over estimated cardinality. The ``Runtime Prediction'' stage in Figure~\ref{fig:system} shows an example of this process, where an initially unlabeled (uncolored) operator is fed through the embedding network to produce an embedded vector, which is subsequently classified by the traditional ML model. The result of this classification can then be utilized by the DBMS, e.g. if the specialized model predicts that a particular operator's cardinality has been underestimated, the DBMS could perform sampling, or replace the particular operator with one less sensitive to cardinality overestimations (e.g. replace a loop join with a hash join).


%% file: training.tex
\section{Operator Embeddings Training}
\label{sec:training}

In this section, we give a technical description of the embedding learning, task-based model training, and runtime predictions phases of the framework we discussed above.

\subsection{Embedding Learning}

The first phase of our framework involves the  training of the embedding network. During this step, we train a neural network to emit operator embeddings using a large history of previously-executed queries. Let us assume a sample query workload $W$ from the target database. We treat each query operator $x \in W$ as a large, sparse vector containing information about $x$ such as the operator type, join algorithm, index used, table scanned, column predicates, etc. Categorical variables are one-hot encoded, and vector entries corresponding to properties that a certain operator does not have are set to zero, e.g. for table scan operators, the vector entry for ``join type'' is set to zero.


We  train our embedding network to predict the children, $C_1(x)$ and $C_2(x)$, of a given query operator $x \in W$.  If a query operator has no children, we define both $C_1(x)$ and $C_2(x)$ as the zero vector, denoted $\vec{0}$, and if a query operator has only one child, we define $C_2(x)$ to be $\vec{0}$ and $C_1(x)$ to be that child. Query operators with more than two children are uncommon, and can either be ignored or truncated (e.g., ignore the 3rd child onward).

\begin{figure}
  \centering
  \includegraphics[width=0.32\textwidth]{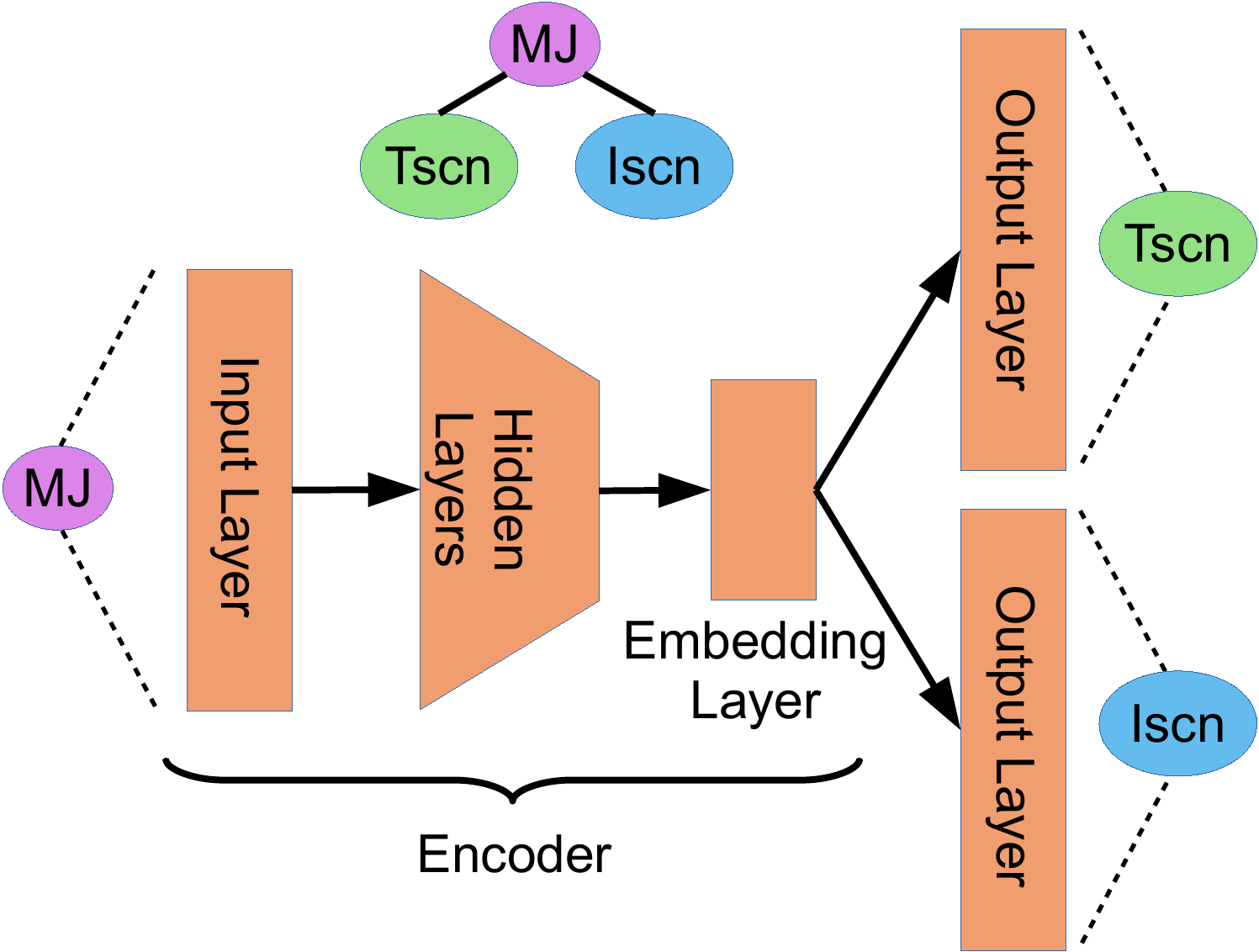}
  \caption{Training the embedding}
  \label{fig:trainingoperator}
\end{figure}

Given a sample workload $W$, we next construct an hour-glass shaped neural network, as shown in Figure~\ref{fig:trainingoperator},  which we train to learn a mapping from each parent to its children. The initial input layer for the embedding network is as large as the sparse encoding of each query operator. The subsequent layers, known as hidden layers, map the output of the previous layer into smaller and smaller layers, corresponding with lower and lower dimensional vectors. The penultimate layer, the \emph{embedding layer}, is the smallest layer in the network. Collectively, these layers are referred to as the \emph{encoder}. Immediately following the embedding layer are the output layers, which attempt to map the dense, compact representation outputted by the embedding layer to the prediction targets, i.e. the children of the input operator.

We refer to the encoder as a function $Enc_\theta(\cdot)$ parameterized by the network weights $\theta$, and we refer to $P^1_\theta(\cdot)$ and $P^2_\theta(\cdot)$ as the functions representing the output layers for predicting the first and second child of the input operator from the output of the encoder, respectively. Thus, the embedding network is trained to by adjusting the weights $\theta$ to minimize the network's prediction error: 

\begin{equation*}
\min_\theta \sum_{x \in W} E(P^1_\theta(Enc_\theta(x)), C_1(x)) + E(P^2_\theta(Enc_\theta(x)), C_2(x))
\end{equation*}

\noindent where $E(\cdot, \cdot)$ represents an error criteria. For vector elements representing scalar variables (such as cardinality and resource usage estimates), we use mean square error, e.g. $E(x, y) = (x - y)^2$. For vector elements representing categorical variables (such as join algorithm type, or hash function choice), we use cross entropy loss~\cite{deep_learning}.
 
By minimzing this loss function via stochastic gradient descent~\cite{sgd}, we train the neural network to accurately predict the contextual information (children) of each input operator. It is important to note that the network will never achieve very high accuracy -- in fact, it is quite likely that the same operator has different children in different queries within the sample workload, thus making perfect accuracy impossible. When this is the case, the best the network can do is match the distribution of the operator's children, e.g. to predict the average of the correct outputs (as this will minimize the loss function), which still requires that the narrow learned embedding contain information about the input operator. The point is not for the embedding network to achieve a high accuracy, but for the network to learn a representation of each query operator that carries a semantic information.

Intuitively, the learned representation is information-rich and dense: the narrow embedding layer forces the neural network to make its prediction of the operator's children based on a compact vector, so the network must \emph{densely} encode information in the embedding layer in order to make accurate predictions, and thus minimize the prediction error. Since the network can make accurate predictions about the context of a query operator given the dense encoding, the dense encoding must be information-rich.

\sparagraph{Example} Figure~\ref{fig:trainingoperator} shows an example of how the embedding network is trained for a single operator: a merge join (MJ) with two children, a table scan (Tscn) and an index scan (Iscn). The network is trained so that when the sparse, vectorized representation of the merge join operator is fed into the input layer, the encoder (which contains a sequence of layers with decreasing sizes) maps the sparse vector into a dense, semantically rich output, which is finally fed into the output layers to predict the merge join's children, the table scan and the index scan. Because the final output layers must make their prediction of the input operator's children based only on the small vector produced by the encoder, the output of the encoder must contain semantic contextual information about the input operator. 

\subsection{Task-based Model Training}
After a good encoder $Enc_\theta$ has been trained, a user can identify a task (e.g. cardinality estimation error prediction) and build a small training set of labeled operators. For the cardinality estimation error prediction task, these labels would indicate whether or not an operator's cardinality was under or over estimated.  We denote each operator in the training set at $x \in T$, with the label of $x$ being $Label(x)$.  Using the embedding network, the user can transform each labeled query operator into a labeled vector, e.g. for any $x \in T$, we compute $Enc_\theta(x)$. These input vectors $Enc_\theta(x)$ and their labels $Label(x)$ can be used as a training set for a traditional, off-the-shelf ML model such as a logistic regression or a random forest.

We denote this off-the-shelf model with parameters $\phi$ as $M_\phi$, and note that, in general, $M$ will be trained to minimize {the model's classification error}:
\begin{equation}
\label{eq:specialize}
\min_\phi \sum_{x \in T} E(M_\phi(Enc_\theta(x)), Label(x))
\end{equation}

\begin{figure}
  \centering
  \includegraphics[width=0.44\textwidth]{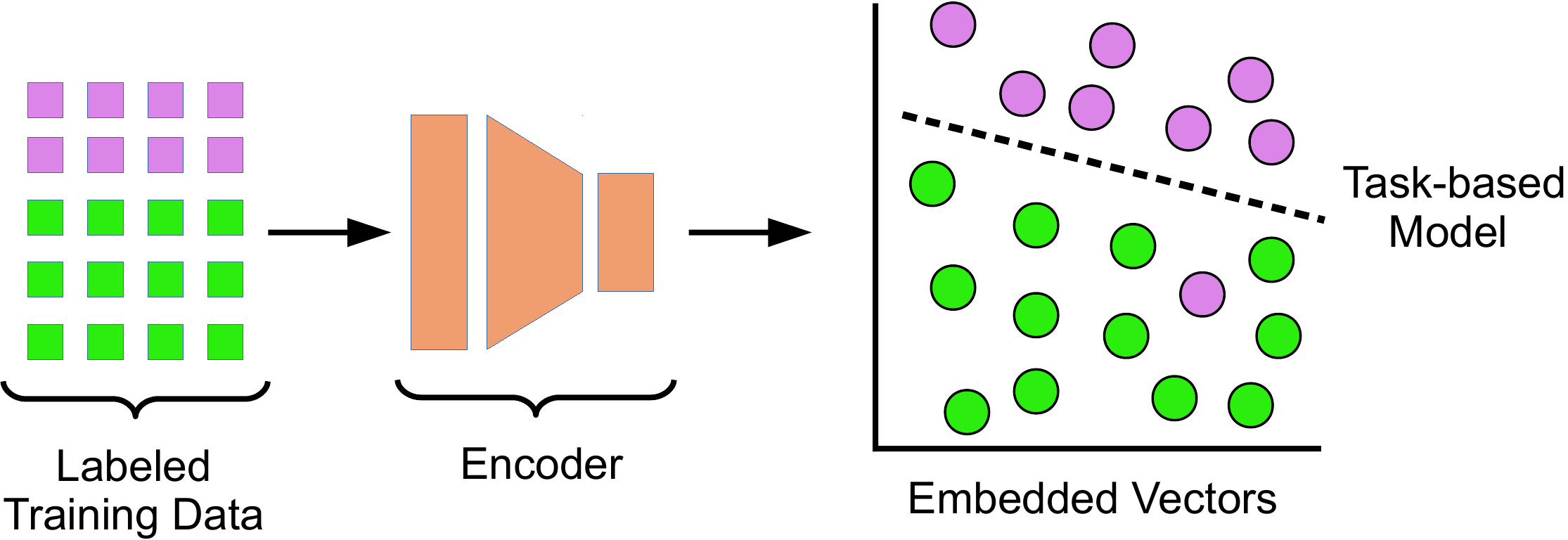}
  \caption{Task-based model}
  \label{fig:specialize}
\end{figure}

Virtually any machine learning model can be used in place of $M_\phi$, and the corresponding learning algorithm can be used to find a good value for $\phi$ (e.g. random forest tree induction for building a random forest model~\cite{rf}, or gradient descent for finding coefficients for a logistic regression~\cite{logistic}). Because the vectors produced by a well-trained encoder are information-rich and dense, operators with interesting properties (such as those with cardinality estimation errors) may be somewhat linearly separable, allowing for fast, simple models like logistic regression to be used with reasonable accuracy. We demonstrate this experimentally in Section~\ref{sec:experiments}.

\noindent{\bf{Training set size}} One extra advantage of using the embedded operator vectors as input to traditional machine learning models is that it reduces the amount of data required to train a effective model compared to training a model directly on labeled sparse operator vectors.  While many deployed database systems have a large log of executed queries available, acquiring \emph{labeled} query operators, i.e. query operators that have been tagged (possibly by hand) with the additional piece of information one wishes to predict, is generally more difficult than analyzing logs. For example, for the cardinality estimation error prediction task, acquiring a large number of query plans from logs is straight-forward, but determining whether the optimizer under or over-estimated the cardinality for each operator requires re-executing the query and recording the estimated and actual cardinalities. Clearly, acquiring this information by re-executing the entire query log is untenable. 

By training the embedding network to predict the context of a query operator using the large supply of easily-acquired unlabeled data, we ensure that the embedded operator vectors contain information about patterns in the query workload. Using these embedded vectors to train a traditional machine learning  model removes the need for the traditional model to re-learn workload-level information, and can thereby achieve strong performance without a large supply of labeled training data.

\sparagraph{Example} An example of task-specific model training for the cardinality error estimation task is depicted in Figure~\ref{fig:specialize}. Here, a small set of training data is represented by squares in the left side of the figure. The color corresponds to the prediction target: whether or not the cardinality estimate for the operator was too high or too low. The training set is then fed through the learned embedding. The scatterplot on the right side of Figure~\ref{fig:specialize} depicts the resulting dense, information-rich vectors (in this case, of dimension 2). A resulting classifier can be trained to find a simple linear boundary (e.g., logistic regression, perceptron, SVM) which nearly-perfectly separates the two classes.

\subsection{Runtime Prediction}
At runtime, an unlabeled operator $x$ can be ran through the encoder $Enc_\theta(x)$, then through the traditional ML model, $M_\phi(Enc_\theta(x))$, in order to produce a predicted label for $x$. The resulting classification decision can be used by the DBMS. For example, for the cardinality error estimation task, if the model $M_\phi$ predicts that an operator $x$'s cardinality has been underestimated, the DBMS could perform additional sampling, refuse to run the query, or ask the user for confirmation. In a more advanced setting, the model's prediction could be combined with a rule-based system to avoid catastrophic query plans, e.g. if the model predicts that a loop join operator has an underestimated cardinality, replace that loop join with a hash join.

We note that because this inference is happening during query processing, inference time matters. As a result, users may wish to select a model with sufficently fast inference time for their application: if model inference time is too high, the net effect on the system may be negative, even if the model provided accurate predictions. We experimentally analyze the inference time required by the encoder, and the encoder combined with various off-the-shelf machine learning models, in Section~\ref{sec:inference_time}.


%% file: experiments.tex
\section{Experimental Analysis}
\label{sec:experiments}

Here, we present an experimental analysis of the operator embedding framework. First, we analyze the embeddings themselves, investigating the properties of the learned vector space that operators are embedded into. This analysis allow us to build an intuition for why the learned embedding approach offers effective feature vectors for use with traditional machine learning algorithms. Then, we measure the effectiveness and efficiency of flexible operator embeddings for several data management tasks.

\sparagraph{Neural network setup} Unless otherwise stated, our hourglass embedding network generates embeddings of size 32, meaning that query operators are mapped into 32 floating-point numbers. This neural network has six hidden layers (of size 256, 256, 128, 128, 64, and 64 nodes). Layer normalization~\cite{layer_norm} (a technique to stabilize the training of neural networks) and ReLUs~\cite{relu} (an activation function) are used after all layers except for the output layer. The network was trained using stochastic gradient descent~\cite{sgd} over 100 epochs (passes over the training data).
The embeddings were trained using a GeForce GTX TITAN GPU and the PyTorch~\cite{pytorch} deep learning toolkit.

\sparagraph{Database setup} Unless otherwise stated, all queries are executed using PostgreSQL 10.5~\cite{url-postgres}, running on Linux kernel 4.18. PostgreSQL was ran from inside a virtual machine with 8 GB of RAM, a configured buffer size of 6GB, and two virtualized CPU cores. The underlying machine has 64GB of RAM and a Intel Xeon E5-2640 v4. 

\sparagraph{Input vectors setup} Queries are initially encoded into a sparse representation based on features we extracted from the output of PostgreSQL's \texttt{EXPLAIN} functionality. The number of features (i.e., the size of the input sparse vectors) depends on the dataset, as different datasets have different numbers of relations, attributes, etc. and thus have different sized sparse input vectors. These features vary between 280 and 478, and contain information such as the optimizer's estimated cost (for all operators) and the expected number of hash buckets required (for hashing operators). Details about the initial sparse encoding can be found in Appendix~\ref{apx:encoding}.

\sparagraph{Dataset} We conducted our experiments over one synthetic and two real-world datasets, which were provided by a large corporation on the condition of anonymity:
\begin{itemize} 
\item{{\bf TPC-DS Workload}: this workload includes 21,688 TPC-DS query instances generated from the 70 TPC-DS templates that are compatible with PostgreSQL without modification. These queries were executed on a TPC-DS dataset with scale factor of 10, resulting in a total database size of 25GB. We have made execution traces of these queries publicly available\footnote{\texttt{http://cs.brandeis.edu/\~{}rcmarcus/tpcds.tar.xz}} for replication and analysis.}
\item{{\bf Online Workload}: we also used a real-world, online workload extracted from execution logs of large corporation. The dataset contains 8,000 analytic (OLAP) queries sent by 4 different users in a 48-hour period. The total database size is 5TB, and the average query reads approximately 300GB of data.}
\item{{\bf Batch Workload}: finally we used a real-world, batch workload executed weekly at a large corporation for report generation. The dataset contains 1,500 analytic (OLAP) queries sent by 98 different users. The total database size is 2.5TB, and the average query reads approximately 350GB of data.}
\end{itemize}

\subsection{Analysis of  Operator Embeddings}

\begin{figure*}
  \centering
  \begin{subfigure}{0.3\textwidth}
    \includegraphics[width=\textwidth]{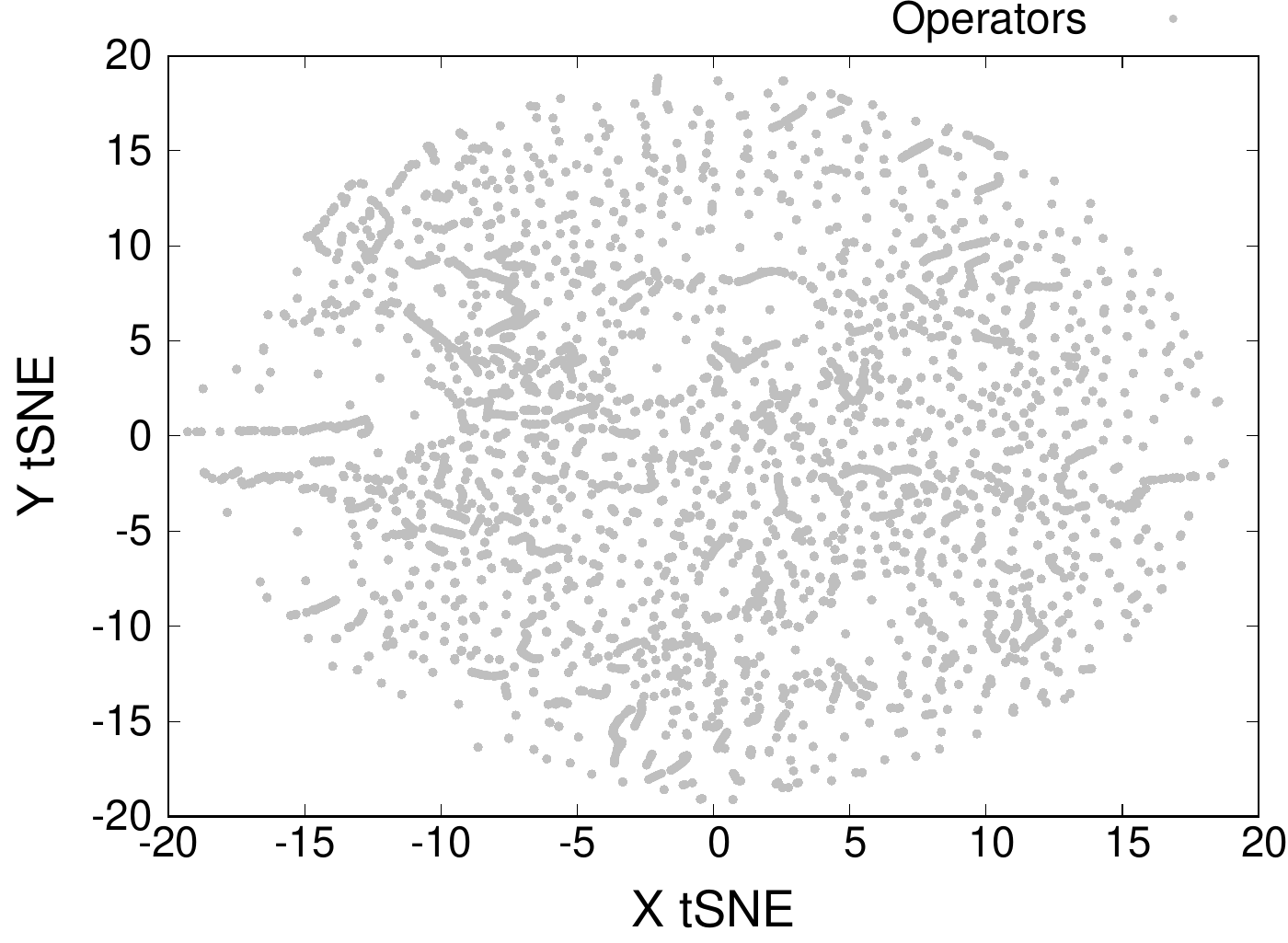}
    \caption{Embedding under t-SNE}
    \label{fig:raw_embedding}
  \end{subfigure}
  \begin{subfigure}{0.30\textwidth}
    \includegraphics[width=\textwidth]{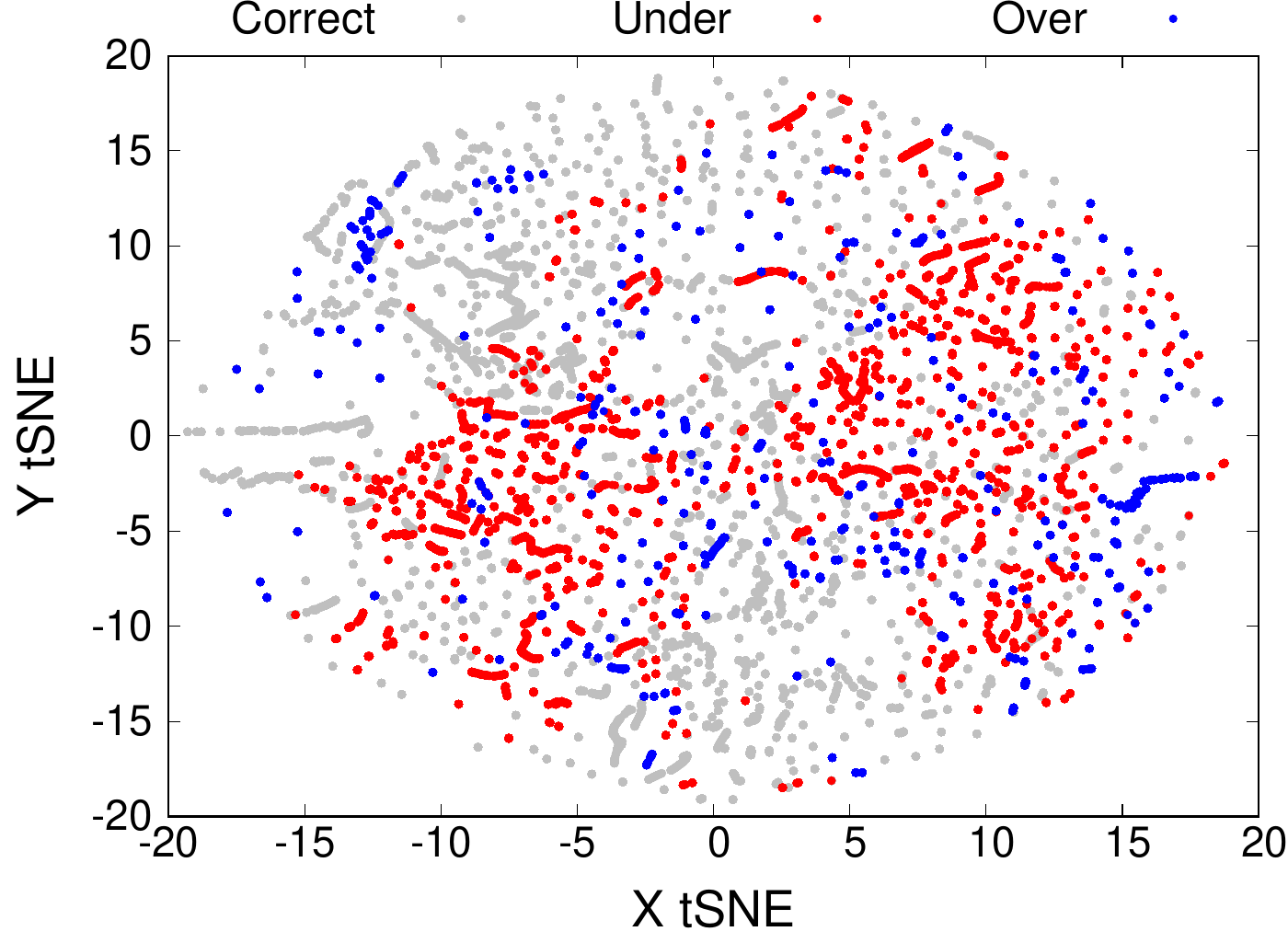}
    \caption{Cardinality errors}
    \label{fig:card_errors}
  \end{subfigure}
  \begin{subfigure}{0.30\textwidth}
    \includegraphics[width=\textwidth]{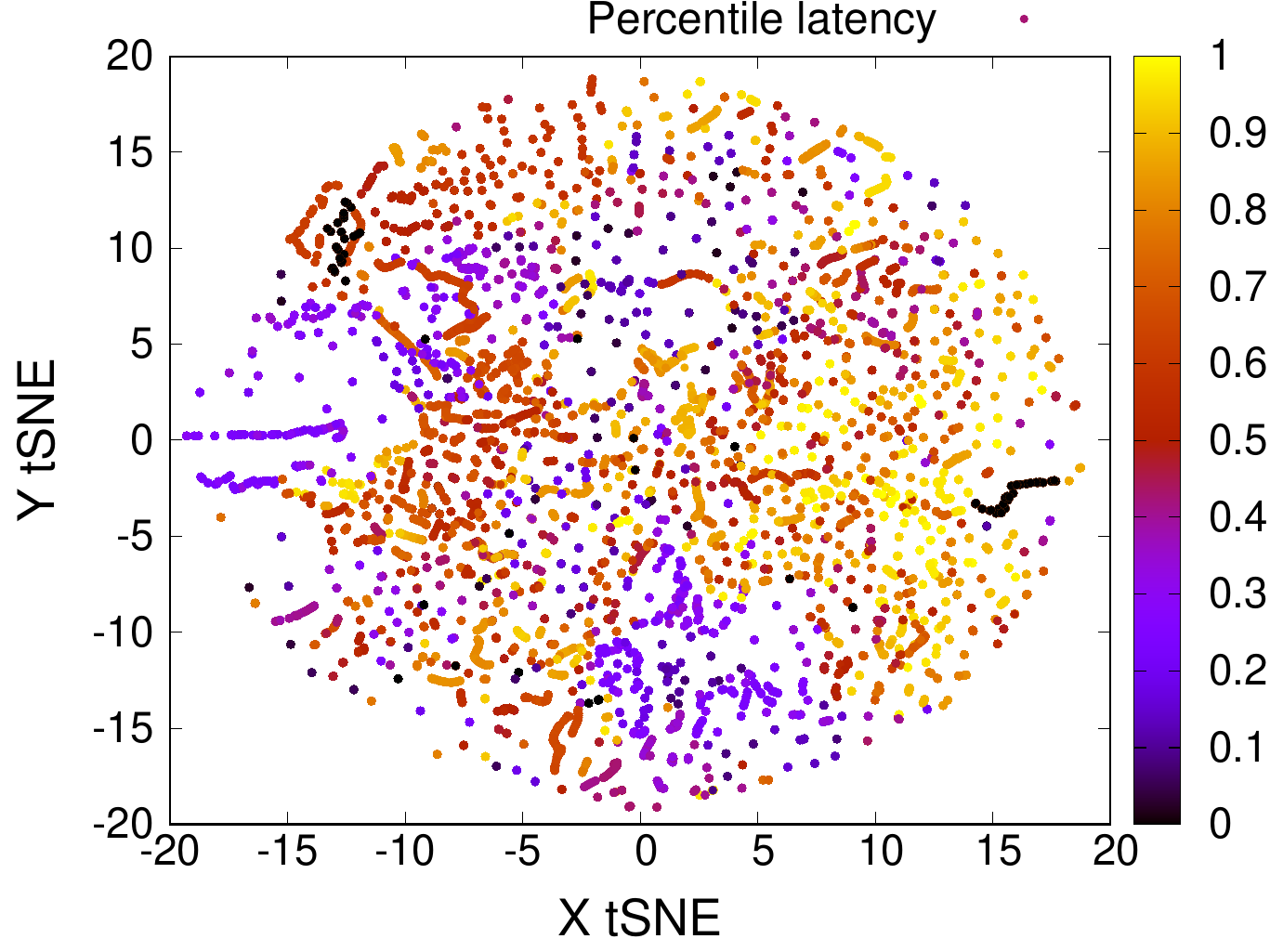}
    \caption{Query latency}
    \label{fig:latency}
  \end{subfigure}
  \caption{t-SNE plots of a learned embedding for TPC-DS (best viewed in color)}
\end{figure*}

First, we explore and analyze the  embeddings learned by the hourglass neural network. This is the output of the first phase of our framework. Here, we trained the neural network on the query plans we collected from the TPC-DS queries. We then ran all the query operators present in those query plans through the trained embedding network, resulting in one 32-dimensional vector for each operator. We refer to these as the \emph{embedded operators}.

\sparagraph{Visualizing embedded operators} To visualize the embedded operators, we used the t-Distributed Stochastic Neighbor Embedding (t-SNE)~\cite{tsne} technique. The t-SNE technique projects high dimensional vector spaces (in our case, the 32-dimensional embedding space), into lower dimensional spaces (2D, for visualization), while striving to keep data that is clustered together in the high dimensional space clustered together in the low dimensional space as well.

Figure~\ref{fig:raw_embedding} shows the t-SNE of the 32-dimensional TPC-DS operator vectors. Note that the X and Y values of a particular point have no meaning on their own -- the t-SNE only attempts to preserve clusters in the higher dimensional space. On its own, Figure~\ref{fig:raw_embedding} does not give much insight into the shape of the learned embedding. However, several clusters and shapes appear, which we analyze next.

\sparagraph{Groupings in the t-SNE} To confirm that the learned embedding carries information about query operator semantics, we next color each embedded operator  based on the accuracy of PostgreSQL's cardinality estimate. In Figure~\ref{fig:card_errors}, operators for which the optimizer correctly estimated the cardinality within a factor of two are colored gray (Correct). When the cardinality is overestimated by at least a factor of two, the operator is colored blue (Over). When the cardinality is underestimated by at least a factor of two, the operator is colored red (Under).

We make two observations about Figure~\ref{fig:card_errors}.
\begin{enumerate}
\item{Even in the 2D space used for visualization, there are many apparent clusters of cardinality under and over estimations. This demonstrates that the operator embedding -- which was trained with no knowledge of cardinality estimation errors -- still learned an \emph{embedding that preserves semantic information} about the query operators. In other words, by learning an embedding useful for predicting the context (children) of an operator, the neural network learned to embed query operators into a vector space with semantic meaning.}
\item{The fact that operators with cardinality estimation errors are clustered together indicates that a \emph{machine learning model should be able to learn underlying patterns} in the embedded data (e.g., the clusters) and make useful predictions. We investigate this directly in Section~\ref{sec:cboost}.}
\end{enumerate}

We also colored the t-SNE plot by operator latency, as shown in Figure~\ref{fig:latency}. Here, we color each query operator based on the percentile of its latency, so that the fastest query operator, the 0th percentile, is colored with dark blue and the slowest query operator, the 100th percentile, is colored with yellow. When comparing Figure~\ref{fig:card_errors} and Figure~\ref{fig:latency} we can observe, unsurprisingly, that many of the slowest query operators correspond with cardinality underestimations, possibly resulting in spills to disk or suboptimal join orderings. Figure~\ref{fig:latency} demonstrates similar behavior to the previous plot. Long-running query operators tend to be neighbors with other long-running query operators, forming clusters in the 2D visualization. These clusters demonstrate that \emph{the learned embeddings carry semantic information} (e.g., information related to operator latency), which can be taken advantage of by a machine learning model.


\subsection{Operator Embeddings Effectiveness}
Next, we evaluated the task-based model training component of our framework. Here, we trained a number of machine learning models using our learned operator embeddings for a number of different data management tasks, and we evaluate the effectiveness of this models. These models are trained using a labeled set of a query operators for a particular task. To show the flexibility of our approach, we measure the performance of the learned embeddings across three different tasks:

\begin{enumerate}
\item{{\bf Query admission}: the model is trained to predict whether a particular query contains an operator that will take an extraordinary amount of time to execute. The decision is used to admit or reject the query.}
\item{{\bf Cardinality boosting}: the model is trained to predict whether the cardinality estimate of a particular operator's output is too high, too low, or correct.}
\item{{\bf User identification}: the model is trained to identify the user that submitted a particular query.  One application of such a model is to test for outlier queries.}
\end{enumerate}

For each task, we used a number of off-the-shelf machine learning models: (1) logistic regression,  (2) random forest~\cite{rf} (\emph{RF}) with  100 trees, (3) k-nearest neighbors~\cite{knn} (\emph{kNN})  configured to account for 6 neighbors using weighted distance (the best value found after an extensive hyperparamter search), and (4) support vector machines~\cite{svm} (\emph{SVM}). In order to demonstrate that task-based models can be trained with relatively little training data, each model is evaluated using 5-fold cross validation, in which one-fifth of the data is used for training and four-fifths are used for testing at a time (the final number reported is thus the median of 5 runs). For the TPC-DS dataset, cross validation folds are chosen based on query templates, so that the query templates in each training set are distinct from queries in each testing set. For the online workload dataset, training and testing sets are chosen so that  all queries in the training set precede all queries in the testing set (i.e., the training set represents the ``past'', and the test set represents the ``future''). For the batch workload dataset, folds are chosen using uniform random sampling without replacement. 

We  compare the performance of each trained machine learning model when  trained using a number of different input feature vectors. The feature engineering  techniques we used for extracting these vectors are the following:
\begin{enumerate}
\item{{\bf \sparse}: this is the raw, unreduced sparse encoding taken directly from PostgreSQL's \texttt{EXPLAIN} functionality (see Appendix~\ref{apx:encoding} for a listing).}
\item{{\bf \neural}: this is the automatically-learned operator embeddings generated using the ``hourglass'' neural network approach presented here.}
\item{{\bf \pca}: here we use feature vectors from the 32 leading principal components (vectors that explain a high percentage of the variance in the data) of the original sparse vectors. These components are found by performing (automated) principal components analysis~\cite{pca} on the sparse input vectors.}
\item{{\bf \fa}: this is an automatic feature engineering process that uses feature agglomeration~\cite{scikit-learn}, a technique similar to hierarchical clustering. This technique builds up features by combining redundant (measured by correlation) features in the sparse input vectors together until the desired number of features (32) is reached.}
\end{enumerate}

\subsubsection{Query Admission Task}
\label{sec:admit}
\begin{figure*}
  \centering
  \begin{subfigure}{0.3\textwidth}
    \includegraphics[width=\textwidth]{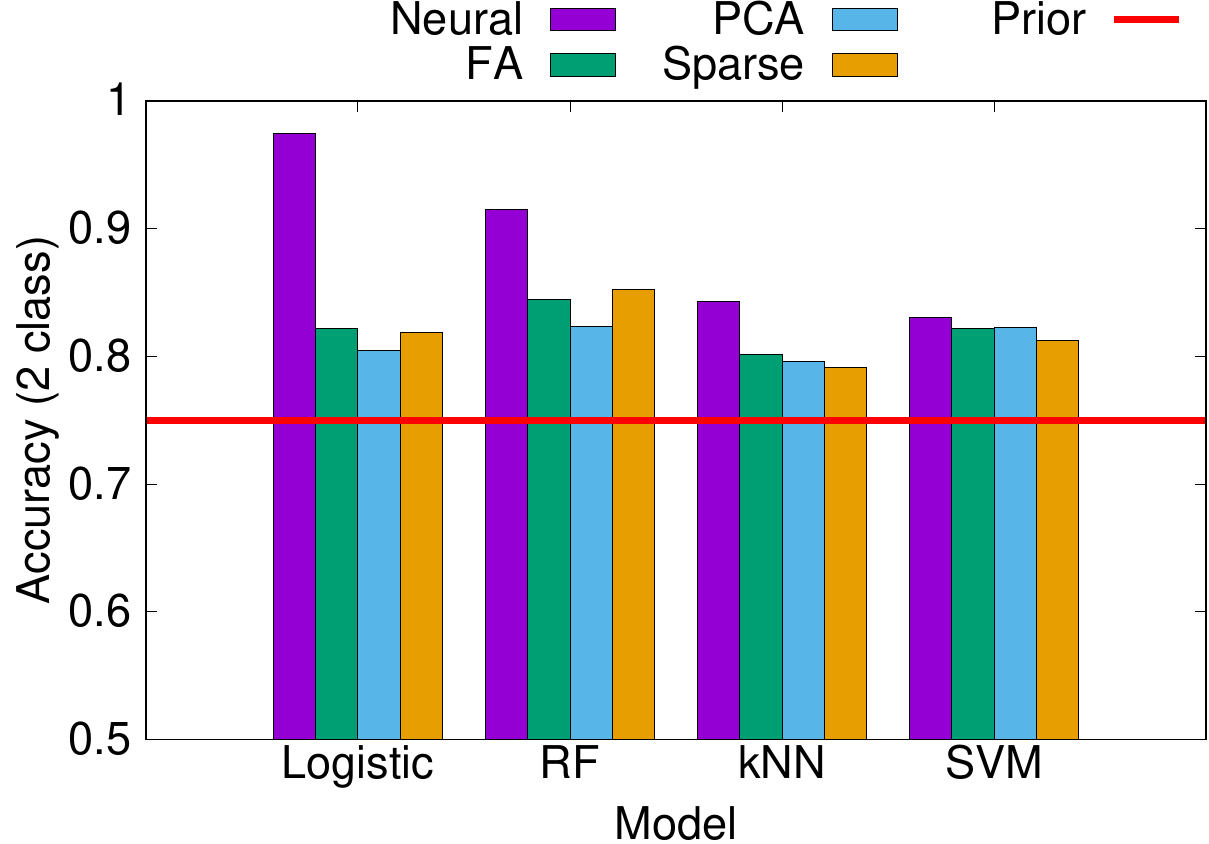}
    \caption{TPC-DS}
    \label{fig:tpcds_admit}
  \end{subfigure}
  \begin{subfigure}{0.30\textwidth}
    \includegraphics[width=\textwidth]{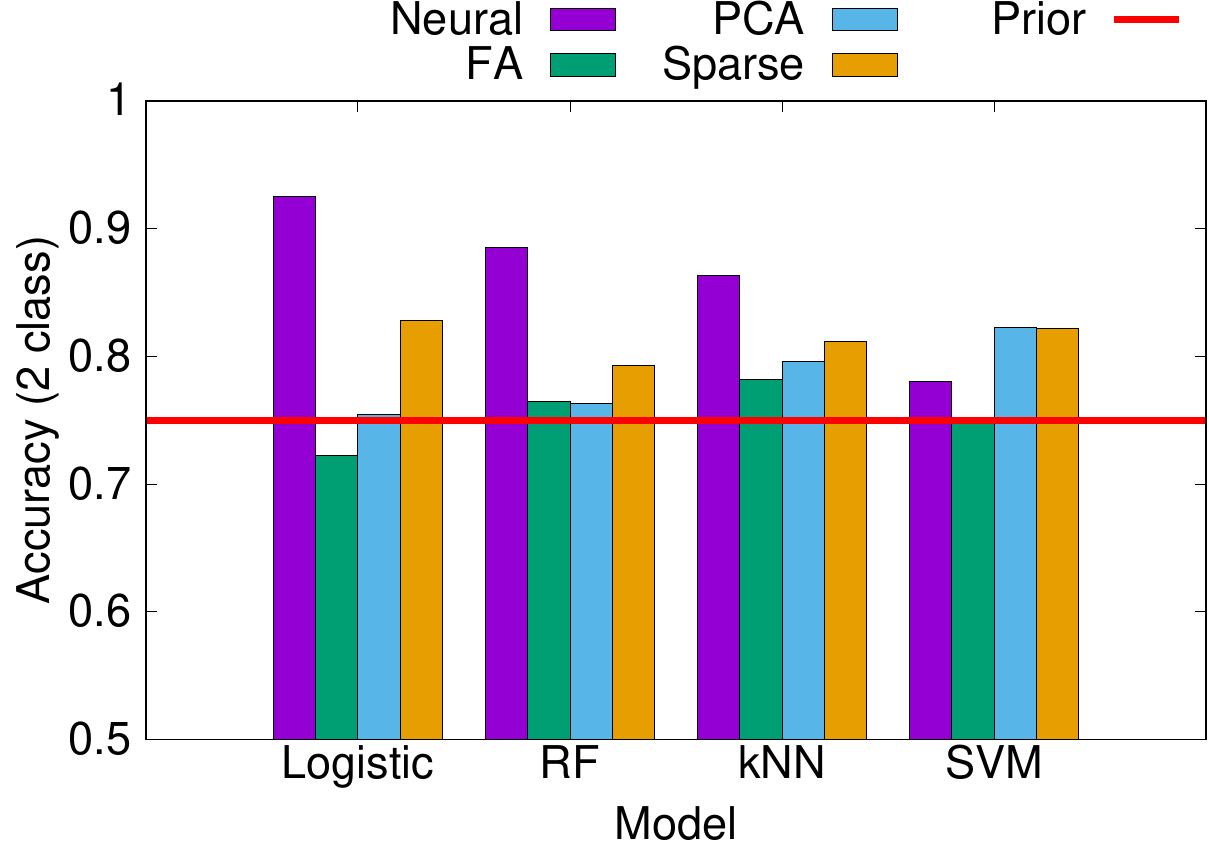}
    \caption{Online Workload}
    \label{fig:ts_admit}
  \end{subfigure}
  \begin{subfigure}{0.30\textwidth}
    \includegraphics[width=\textwidth]{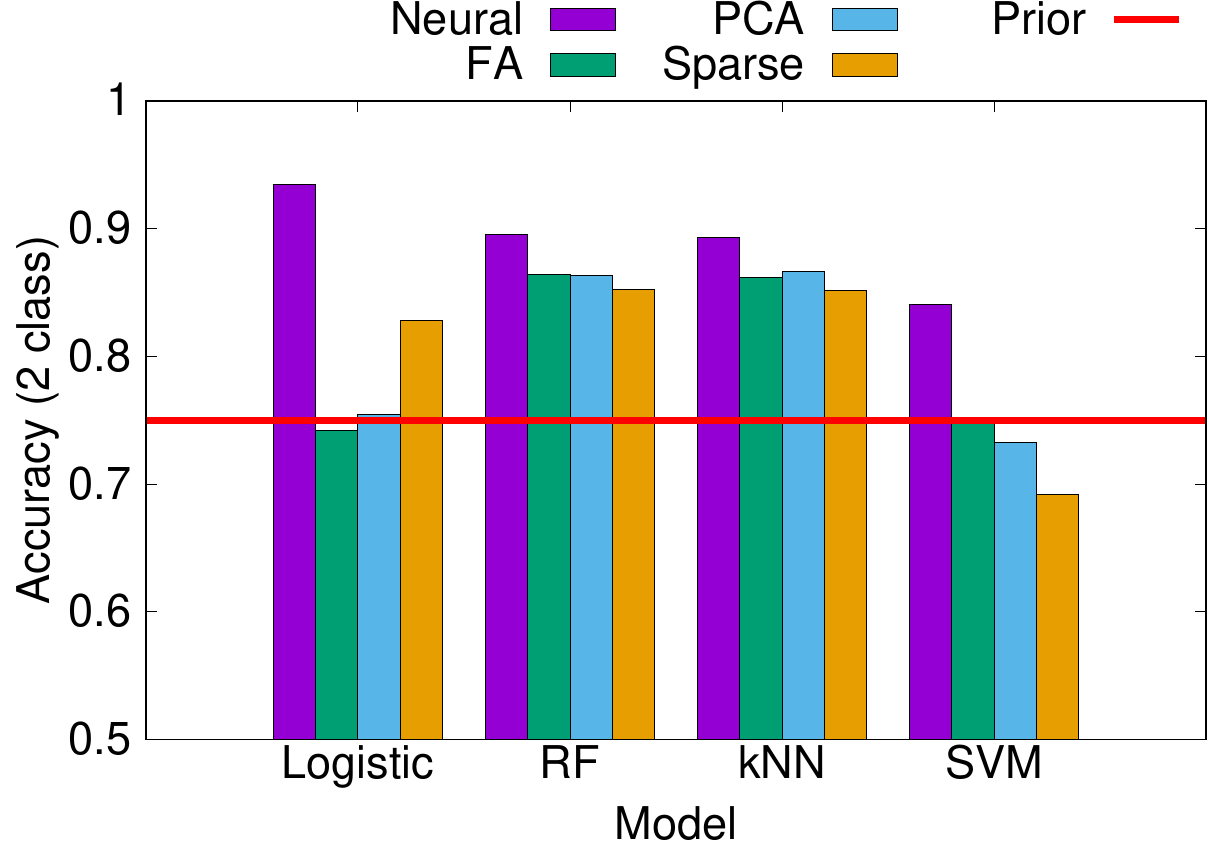}
    \caption{Batch Workload}
    \label{fig:bq_admit}
  \end{subfigure}
  \caption{Query admission prediction accuracy for different models and feature vectors. The models predict if a query does or does not contain an operator with latency above the 95th percentile.}
  \label{fig:admit}
\end{figure*}

For the query admission task, we trained various machine learning models to predict whether or not any operator in a query plan would fall above or below the 95th-percentile for latency. In other words, the model tries to predict if any operator in a given query plan will take longer than 95th\% of previously seen operators or not. To do this, we applied the trained model on \emph{each} operator in an incoming query, and if the model predicts that \emph{any} operator in the query would exceed the 95th percentile threshold, the query is flagged. DBMSes may wish to reject such flagged queries, or prompt the user for confirmation before utilizing a large amount of (potentially shared) resources to execute them~\cite{q-cop, activesla, opt_est_par}. 

Figure~\ref{fig:admit} shows the average 5-fold cross validation accuracy for each model on all three datasets. The red line at $0.75$ represents the prior (i.e., since 75\% of queries do not have an operator exceeding the 95th latency percentile, a model that always guessed this most common class would achieve an accuracy of 75\%).

For all three datasets, the specialized models trained using the \neural input (the learned operator embeddings) outperformed models trained with the other input vectors. While \emph{the combination of the operator embeddings (Neural) and logistic regression saw the highest performance and outperformed other combinations across all three datasets by 10-15\%}, the operator embeddings (Neural) also produced the most effective models compared with any other feature engineering approach. The only exception is the SVM model on the online workload, where the model trained on the \neural inputs was within 5\% of the most effective model.

\sparagraph{Surprising performance of logistic regression} For the query admission task (and the other tasks analyzed next), the combination of the learned embeddings with a logistic regression classifier is surprisingly effective (97\% accuracy for TPC-DS). This is due to the fact that the logistic regression model has a very similar mathematical form to the cross-entropy and mean-squared error loss functions used to train the neural network. Training a logistic regression model on the embedded data is equivalent to re-training the last layer of the embedding network for a different prediction target, a technique called knowledge distillation or transfer learning, which has been shown to be extremely effective~\cite{transfer, transfer2}. This also explains the large gap (15\% - 18\%) between the performance of logistic regression when using the \neural featurization and with the other featurizations.

\subsubsection{Cardinality Boosting Task}\label{sec:cboost}
\begin{figure*}
  \centering
  \begin{subfigure}{0.3\textwidth}
    \includegraphics[width=\textwidth]{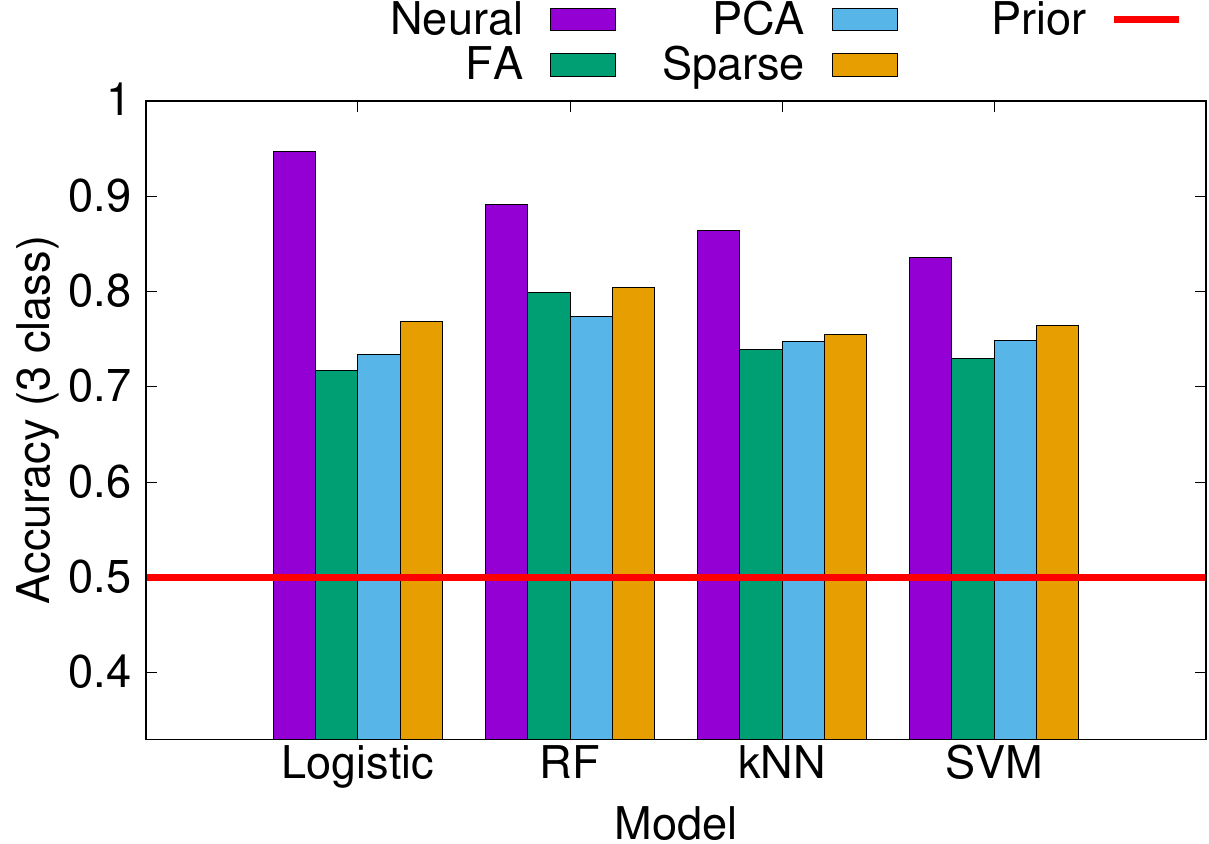}
    \caption{TPC-DS}
    \label{fig:tpcds_boost}
  \end{subfigure}
  \begin{subfigure}{0.30\textwidth}
    \includegraphics[width=\textwidth]{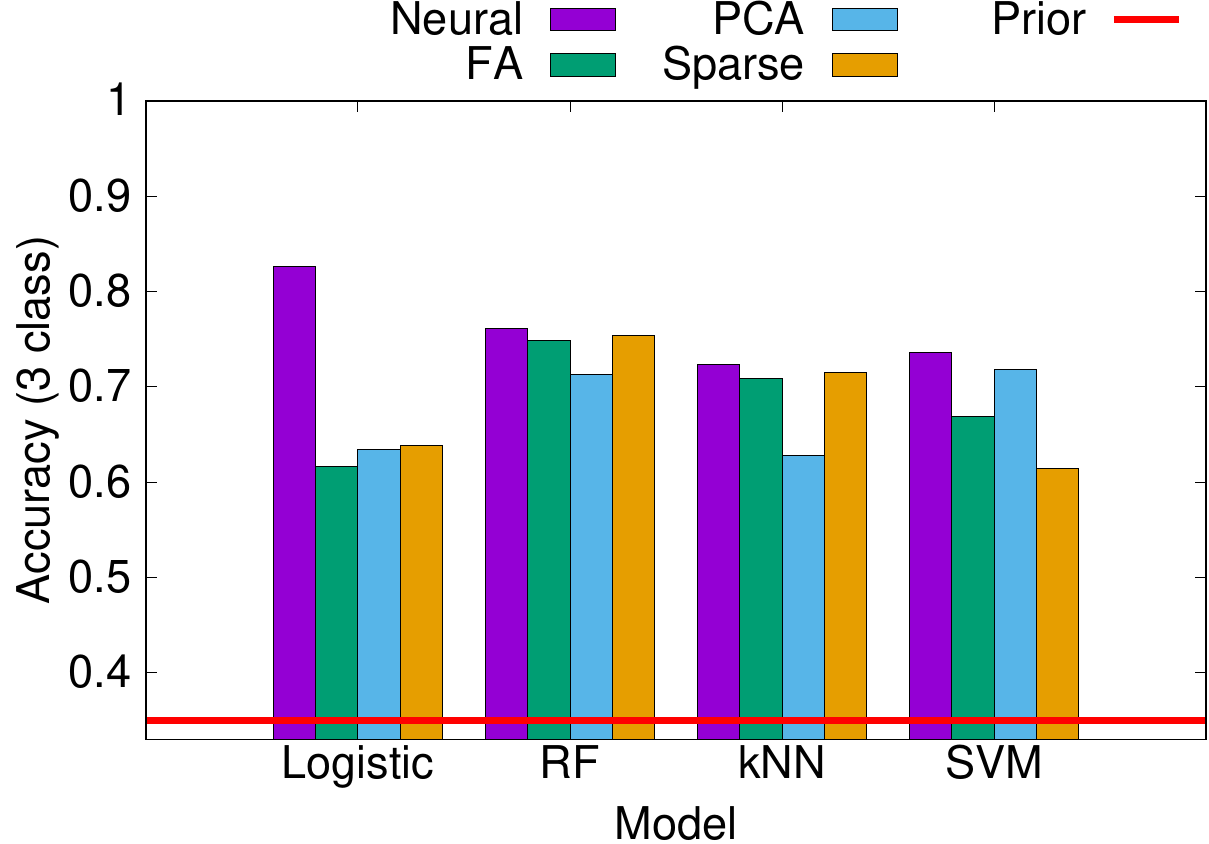}
    \caption{Online Workload}
    \label{fig:ts_boost}
  \end{subfigure}
  \begin{subfigure}{0.30\textwidth}
    \includegraphics[width=\textwidth]{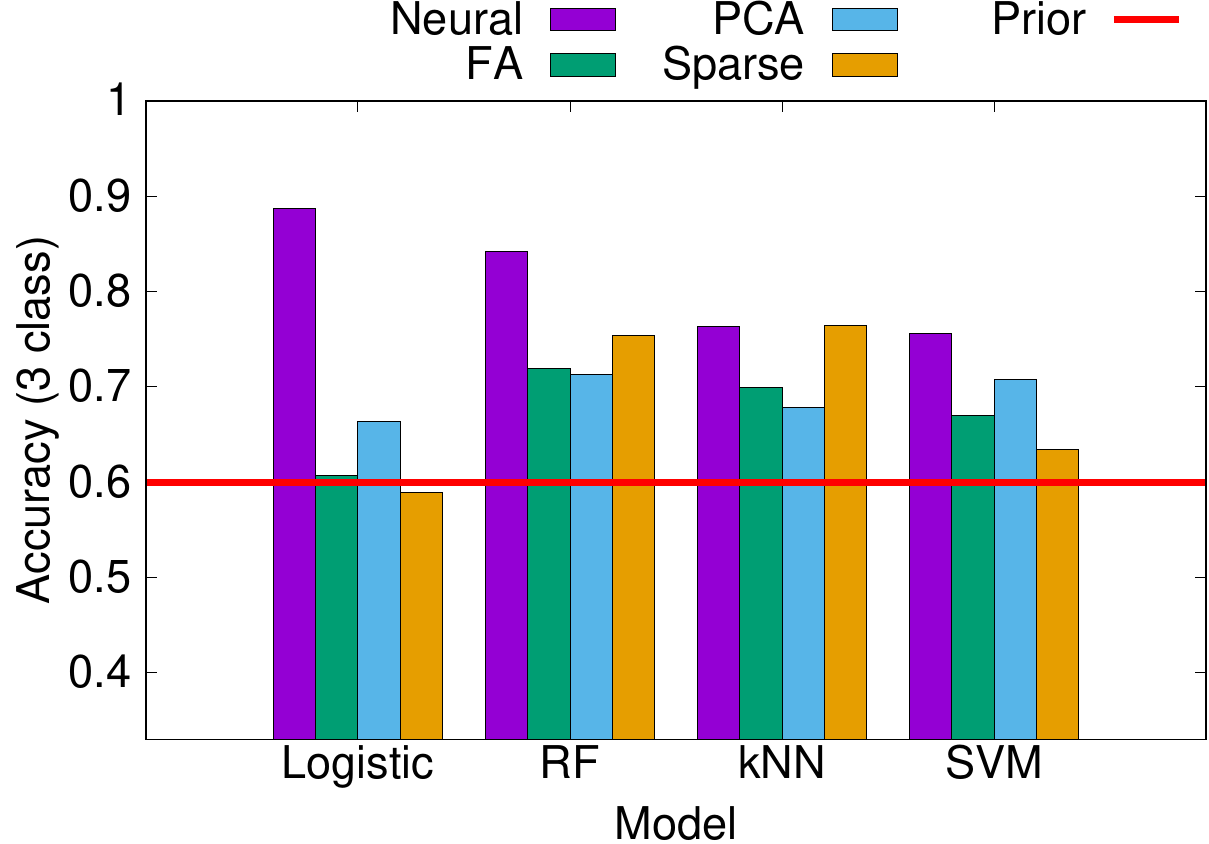}
    \caption{Batch Workload}
    \label{fig:bq_boost}
  \end{subfigure}
  \caption{Cardinality over/under estimation prediction accuracy for different models and feature vectors. The models predict if the optimizer's cardinality estimation is within a factor of two, over by a factor of two, or under by a factor of two.}
  \label{fig:boost}
\end{figure*}

Next, we trained various machine learning models to predict whether or not the PostgreSQL query optimizer's cardinality estimate would be correct, too high (by at least a factor of two), or too low (by at least a factor of two). We call this task cardinality boosting because of the similarities to ``boosting'' techniques in machine learning~\cite{boosting}, i.e. using one model to predict the errors of another model. Providing good cardinality estimates is extremely important for query optimization~\cite{howgood}, and many recent works have applied machine learning to this problem~\cite{deep_card_est, deep_card_est2, qo_state_rep}.

Figure~\ref{fig:boost} shows the performance of the trained machine learning models for the cardinality boosting task across each dataset with each feature engineering technique. Again, \emph{the combination of the \neural featurization and logistic regression produced the most effective model for all three datasets, achieving over 93\% accuracy on the TPC-DS dataset. } While the gap between the \neural featurization and the other featurizations appears significantly higher for the cardinality boosting task than the query admission task, this is attributable to the change in the prior (i.e., 75\% for the query admission task and strictly less than 50\% for the cardinality boasting task).

The strong performance of the \neural featurization is well-explained by the t-SNE plot in Figure~\ref{fig:card_errors}. Since we know the learned embedding is clustering operators with similar cardinality estimation errors together, it is not surprising that machine learning models can find separation boundaries/cluster centers within the data.

\subsubsection{User Identification Task}
\begin{figure*}
  \centering
  \begin{subfigure}{0.3\textwidth}
    \includegraphics[width=\textwidth]{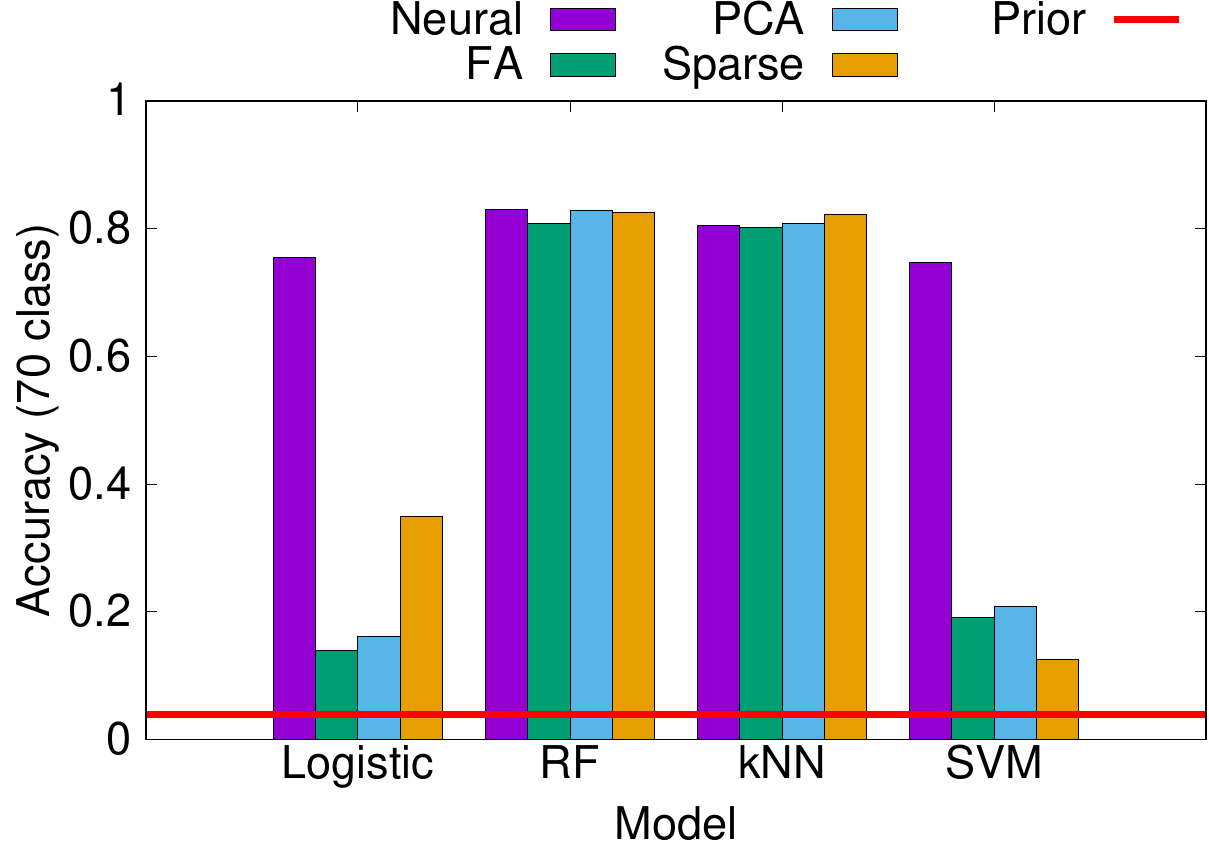}
    \caption{TPC-DS}
    \label{fig:tpcds_user}
  \end{subfigure}
  \begin{subfigure}{0.30\textwidth}
    \includegraphics[width=\textwidth]{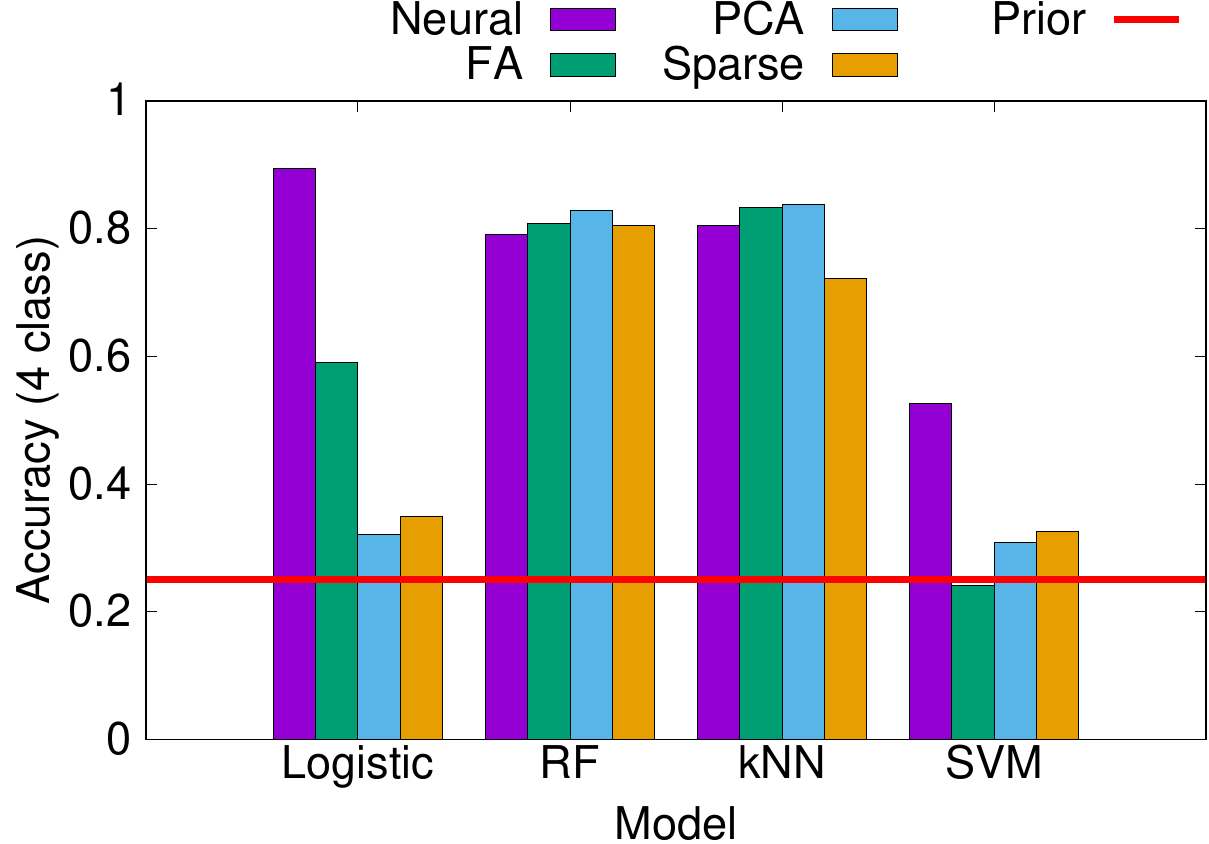}
    \caption{Online Workload}
    \label{fig:ts_user}
  \end{subfigure}
  \begin{subfigure}{0.30\textwidth}
    \includegraphics[width=\textwidth]{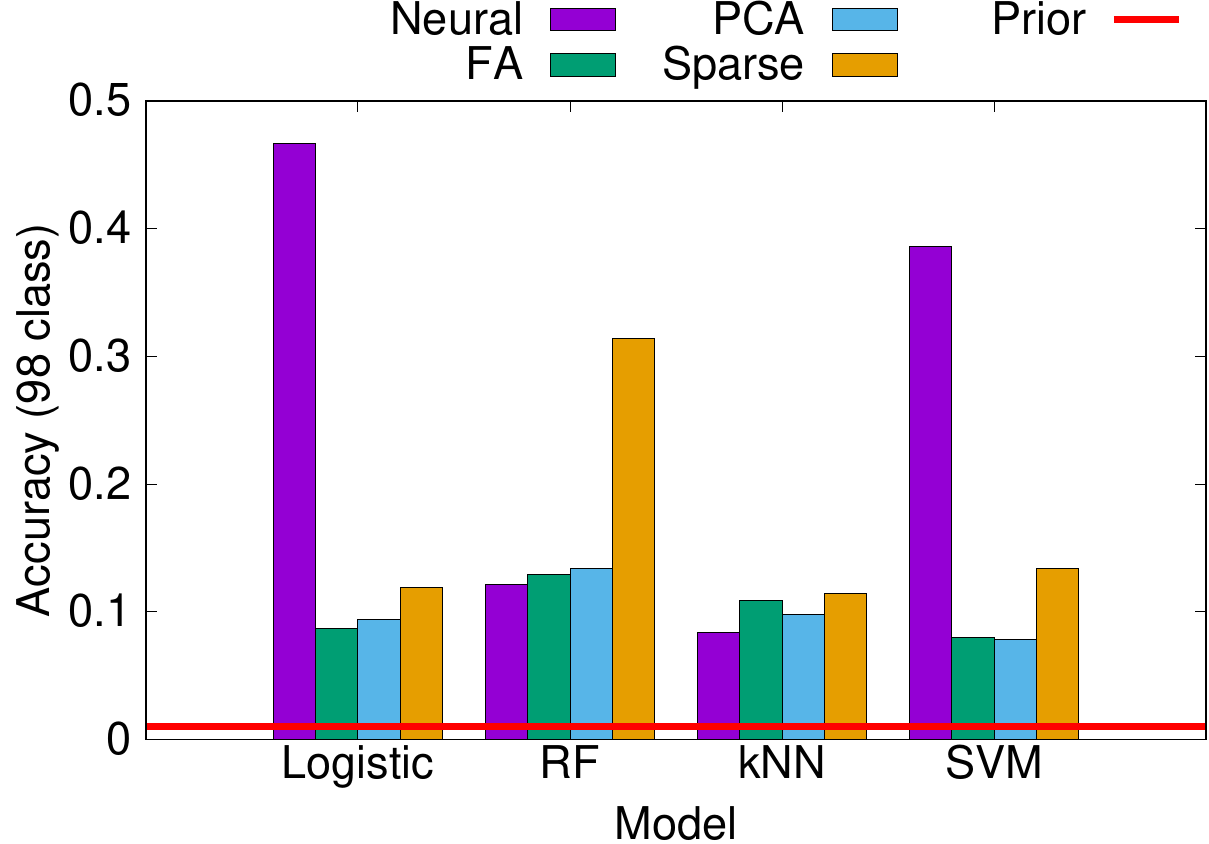}
    \caption{Batch Workload (note y-axis scale)}
    \label{fig:bq_user}
  \end{subfigure}
  \caption{Query user identification prediction accuracy for different models and feature vectors. The specialized model tries to predict the user that sent a particular query (number of classes varies by dataset).}
  \label{fig:user}
\end{figure*}

The last task we evaluate is user identification. In this task, the model's goal is to predict the user who submitted the query containing a particular operator. While the DBMS generally knows the user submitting a query, such a model is useful for determining when a user-submitted query does not match the queries usually submitted by that user, a common learning task in database intrusion detection~\cite{dbsafe, cyberrank, LorenzoBossiSystemProfilingMonitoring2014, Khannewintrusiondetection2007} or query outlier detection~\cite{sql_embed}.

For the TPC-DS data, we use the query template as the ``user'' for a each query (and therefore we use random cross-validation folds). For the online and batch datasets, user information was provided by their respective corporations. The online dataset contains queries submitted from 4 users, whereas the batch dataset contains queries submitted by 98 users. In all cases, the number of queries submitted by each user is approximately equal. Figure~\ref{fig:user} shows the results.

The TPC-DS and batch workloads show an extremely large gap between the logistic regression performance using the \neural featurization and the other featurizations, again because of the significantly lower prior. Overall, \emph{the \neural featurization outperforms all other baselines, demonstrating that the learned embedding contains rich, semantic information about each query operator.}

Two exception, however, are notable: first, for the TPC-DS dataset, the kNN model with the \sparse featurization outperforms the \neural featurization by approximately 1\% (Figure~\ref{fig:tpcds_user}). This is due to certain operators -- such as max aggregates over specific columns -- appearing in only one query template. Such an operator will have a very close neighbor in the sparse vector space, allowing the kNN model to easily classify it. This advantage, however, does not extend to real-world data (e.g. Figures~\ref{fig:ts_user}~and~\ref{fig:bq_user}), where such uniquely identifying operators do not exist.

Second, the random forest algorithm exhibits surprisingly good performance using sparse inputs for the batch workload (Figure~\ref{fig:bq_user}) -- significantly better than the \neural featurization and random forest, although not as good as the \neural featurization and the logistic regression. The reason the \sparse encoding works so well with the random forest model is due to the specifics of the batch dataset: while most users access every table, almost all users can be uniquely identified by the \emph{set} of tables they accessed. The random forest algorithm, which builds a tree of rules based on discrete splitting points, is especially well-suited to identifying the table usage patterns of each user in the one-hot encoding. We note, however, that the logistic regression combined with the \neural feature vectors still outperformed all random forest models on this task.

\subsubsection{Impact of Embedding Size}
\begin{figure*}
  \centering
  \begin{subfigure}{0.3\textwidth}
    \includegraphics[width=\textwidth]{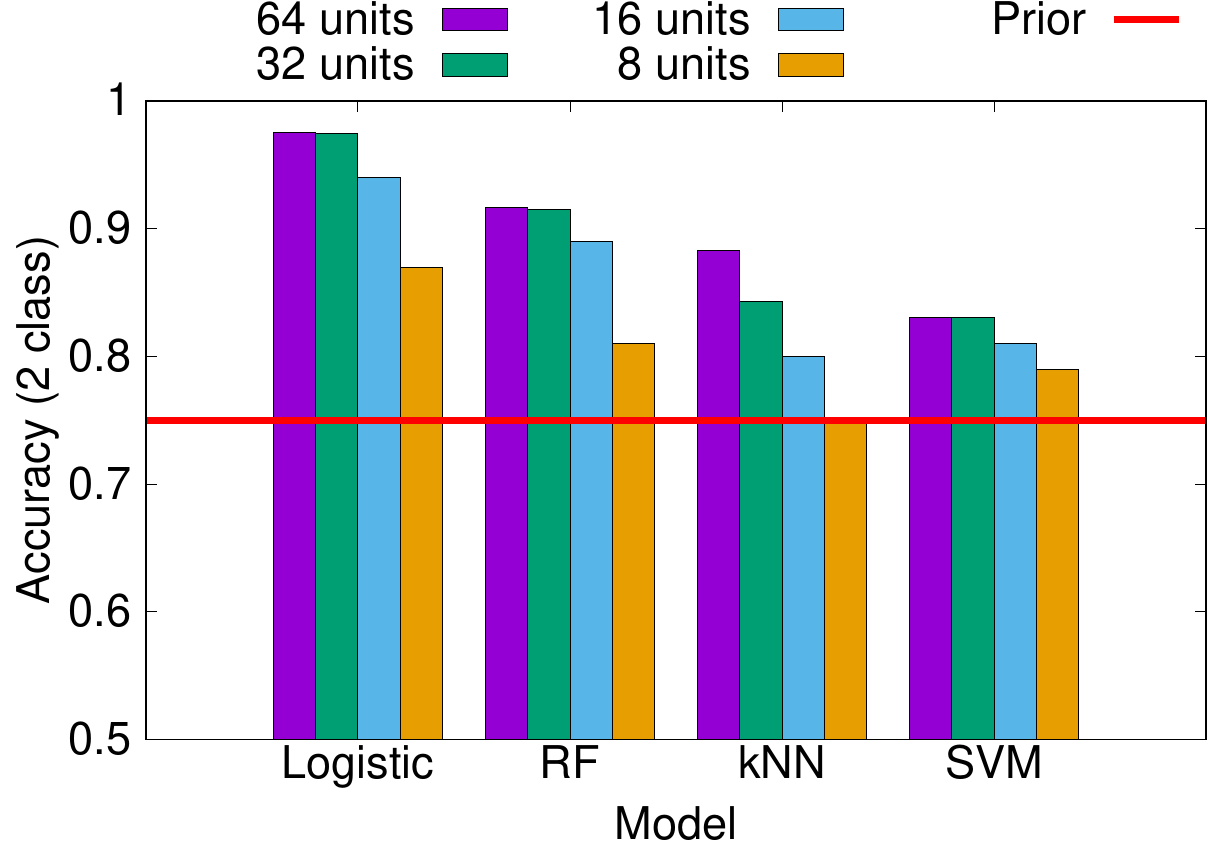}
    \caption{Query Admission}
    \label{fig:es_admit}
  \end{subfigure}
  \begin{subfigure}{0.30\textwidth}
    \includegraphics[width=\textwidth]{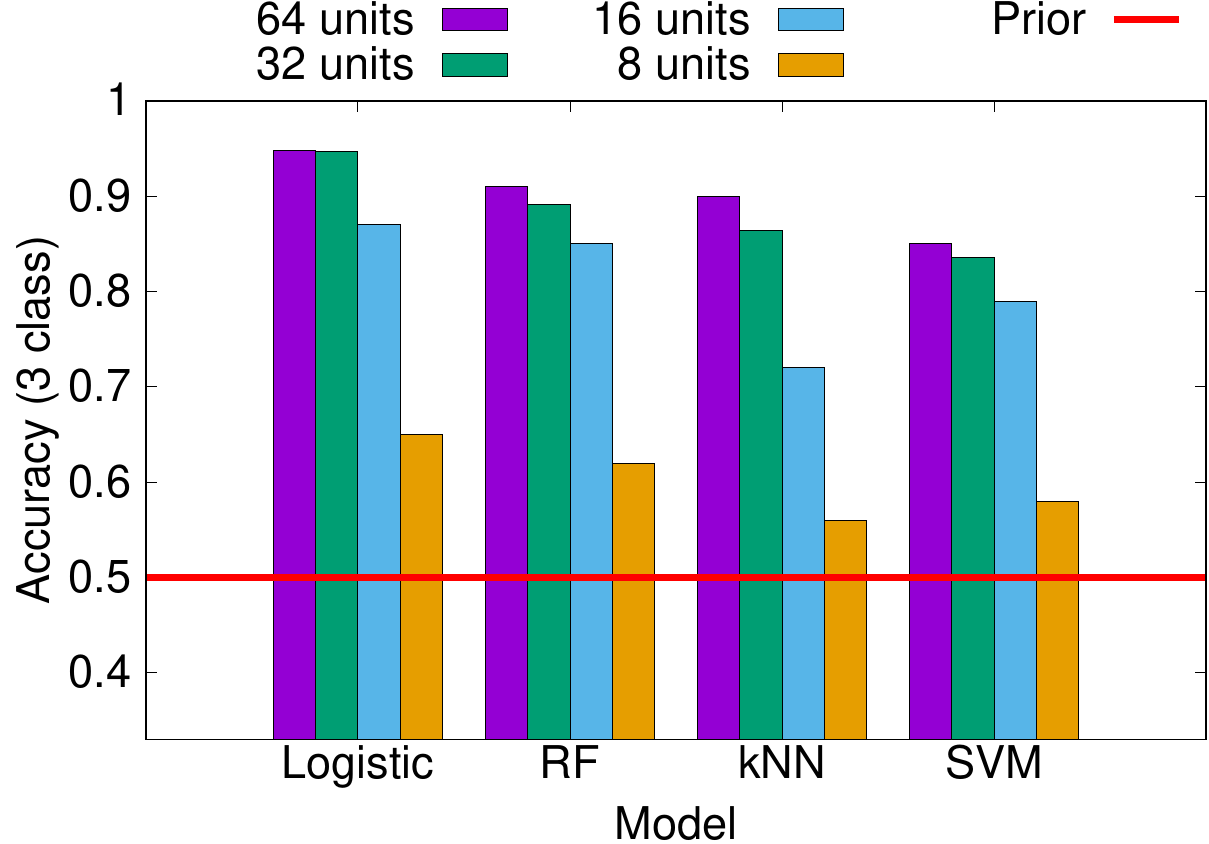}
    \caption{Cardinality Boosting}
    \label{fig:es_boost}
  \end{subfigure}
  \begin{subfigure}{0.30\textwidth}
    \includegraphics[width=\textwidth]{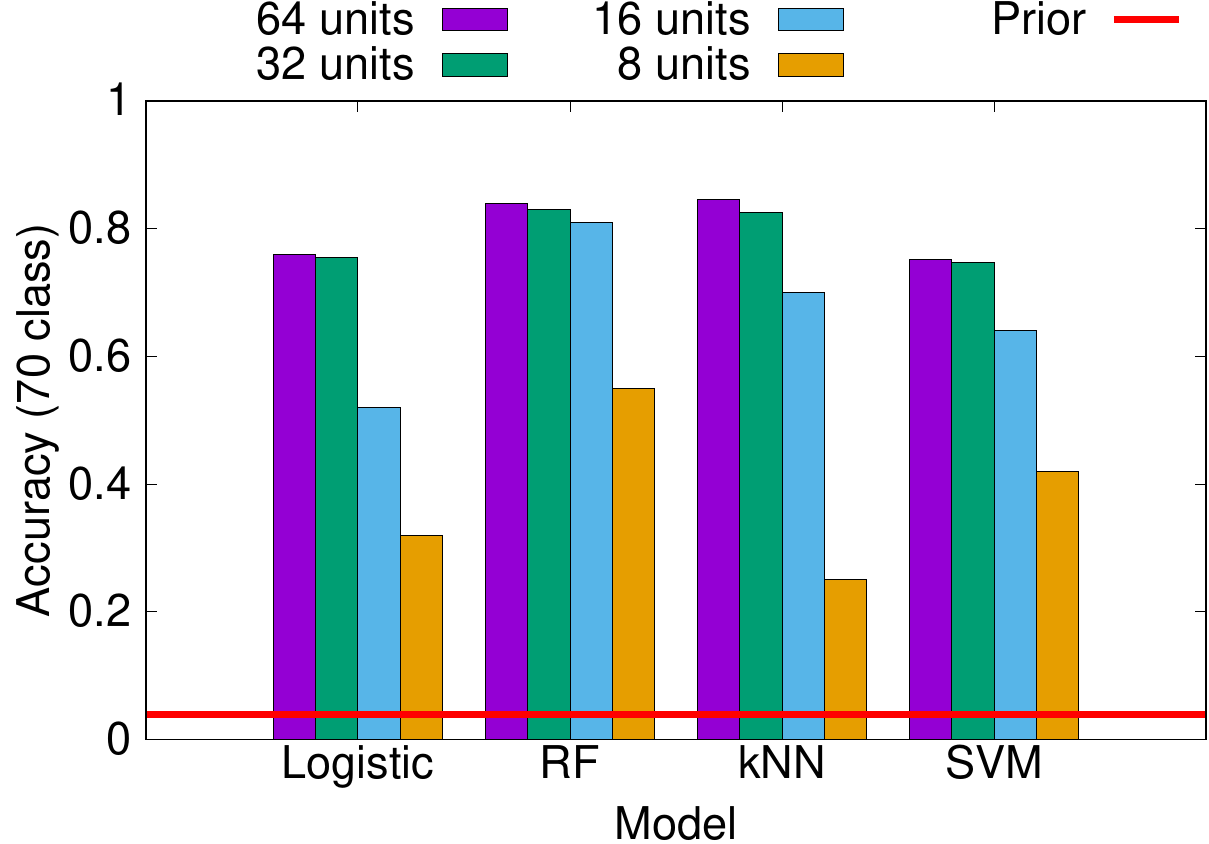}
    \caption{User Identification}
    \label{fig:es_user}
  \end{subfigure}
  \caption{Accuracy of various models on different tasks for the TPC-DS dataset, varying the size of the embedding layer.}
  \label{fig:es}
\end{figure*}

Up until this point, we have only used learned embeddings with embedding layers of size 32, i.e. query operators are mapped into a vector space with 32-dimensions. Here, we evaluate changing this hyperparameter. Figure~\ref{fig:es} shows how the performance of various models change for the TPC-DS dataset when the size of the embedding is set to 64, 32, 16, and 8 dimensions. We observed similar results for both the online and batch workloads, but these plots are omitted due to space constraints.

Generally, the performance of the size-64 embeddings and the size-32 embeddings are nearly identical, indicating that adding additional dimensions beyond 32 to the embedding space does not cause the embedding network to learn a better compressed representation. On the other hand, the performance of all models drops significantly between the size-32 and size-16 embeddings, and again between the size-16 embeddings and size-8 embeddings (best depicted in Figure~\ref{fig:es_user}). This suggests that when the embedding size is smaller than 32, the ``information bottleneck'' (see Section~\ref{sec:training}) becomes too narrow, and the neural network is unable to learn a sufficiently rich description of each query operator in such a small vector.

The ideal embedding size is hard to predict ahead of time, and although our experiments show that an embedding size of 32 or 64 provides good results on a number of datasets, we suggest that users test several configurations. Doing so can be done in an automatic manner by training multiple embedding sizes, and then selecting the one that results in the best cross-validated model performance.

\subsection{Runtime Efficiency}
\label{sec:inference_time}
\begin{figure*}
  \centering
  \begin{subfigure}{0.3\textwidth}
    \includegraphics[width=\textwidth]{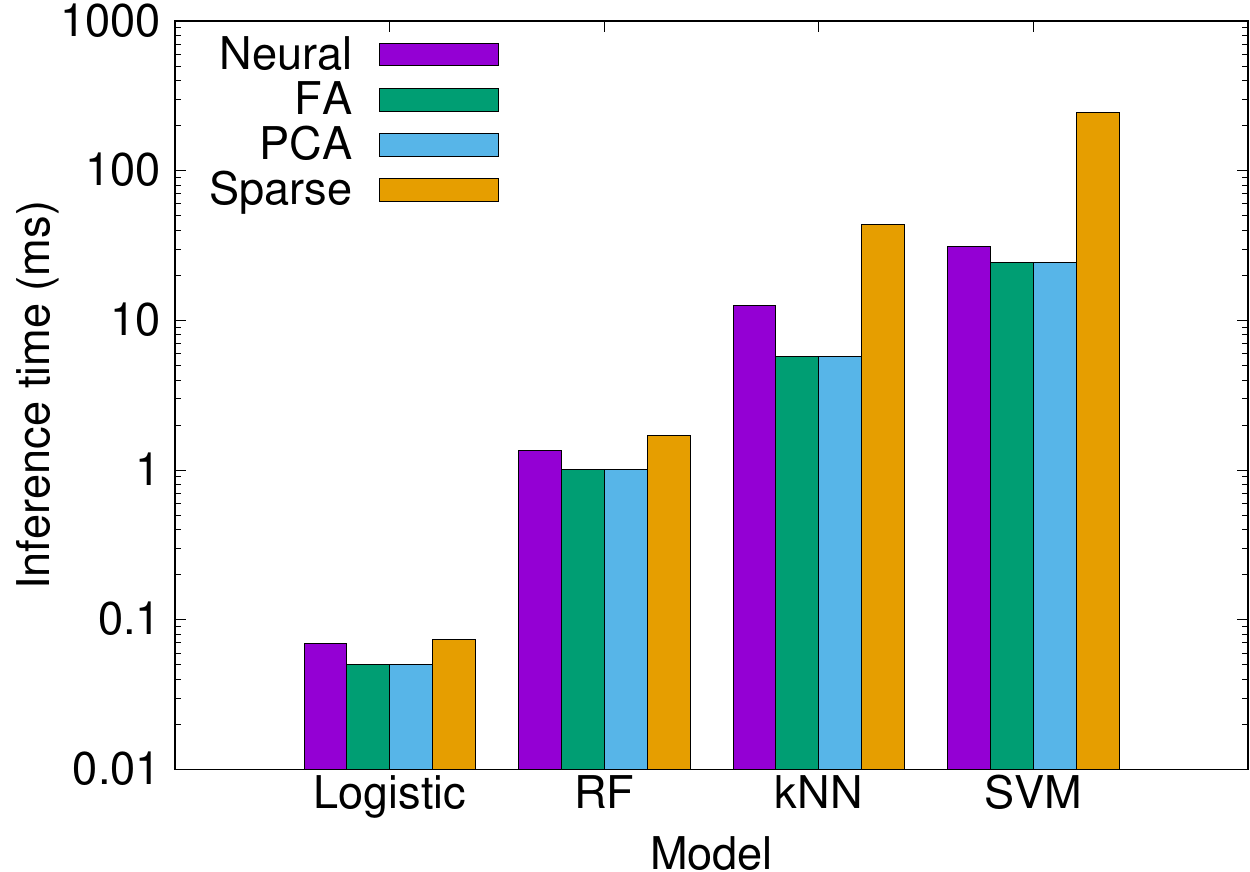}
    \caption{Inference time}
    \label{fig:throughput}
  \end{subfigure}
  \begin{subfigure}{0.3\textwidth}
    \includegraphics[width=\textwidth]{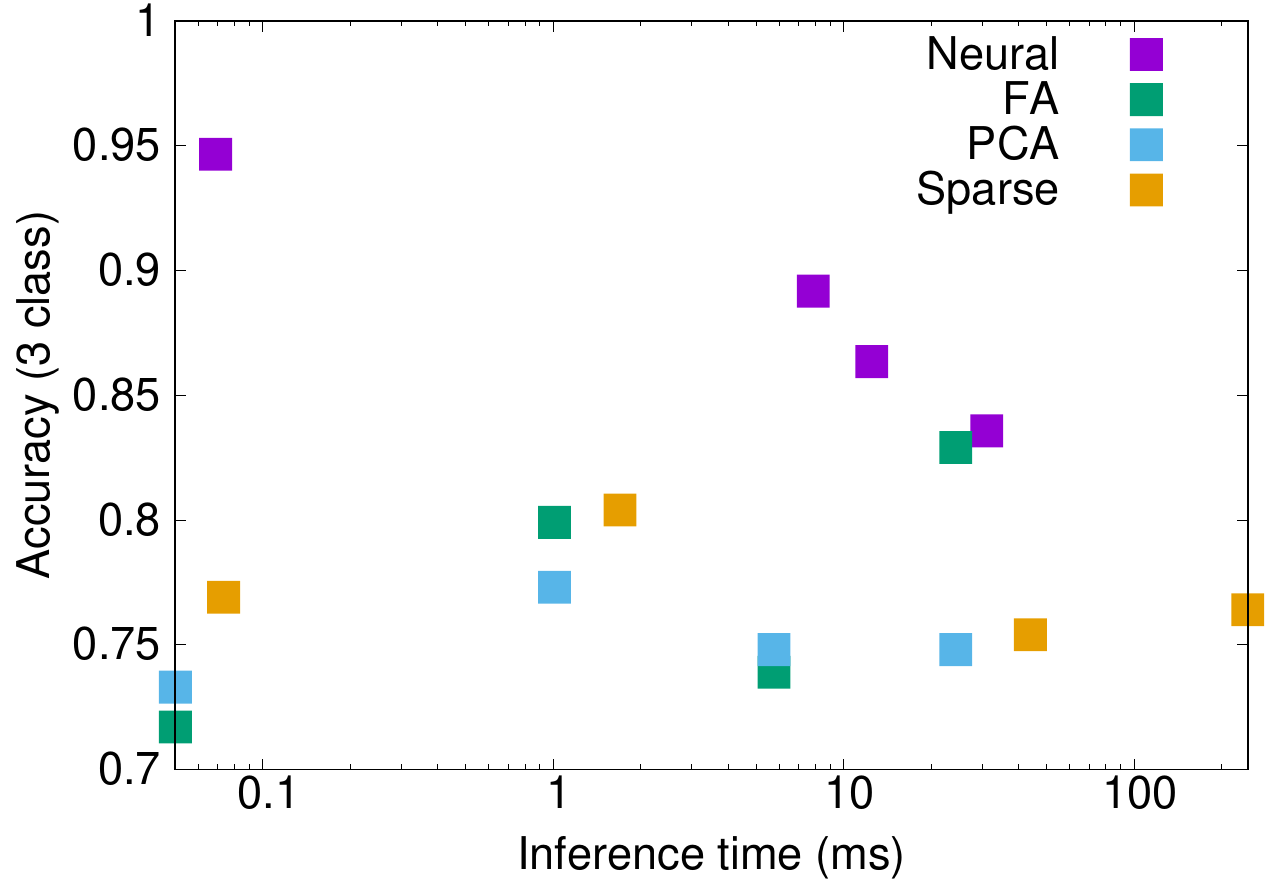}
    \caption{Cardinality boosting, TPC-DS}
    \label{fig:pareto}
  \end{subfigure}
  \begin{subfigure}{0.3\textwidth}
    \includegraphics[width=\textwidth]{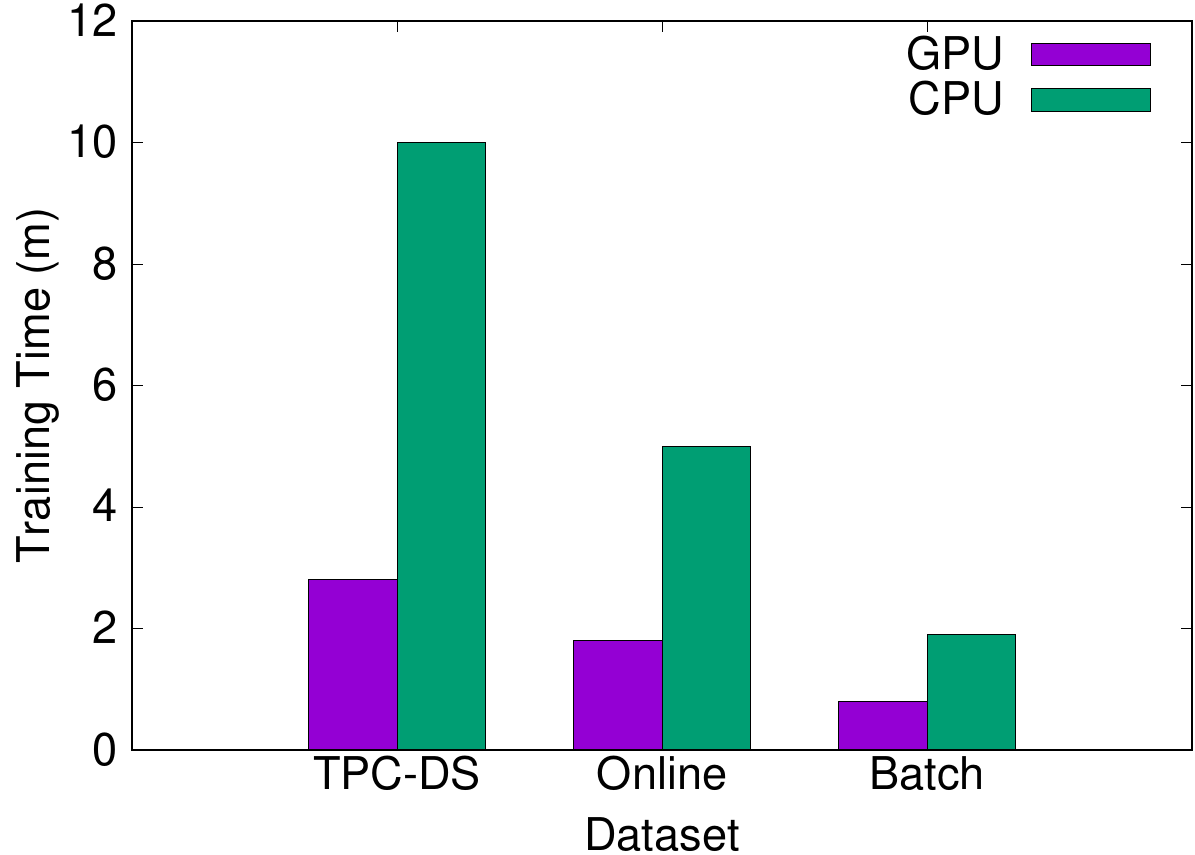}
    \caption{Training time}
    \label{fig:training}
  \end{subfigure}
  \caption{Analysis of inference time, training time, and accuracy}
  \label{fig:runtime}
\end{figure*}

After an embedding has been trained and a subsequent machine learning model has been trained for a particular task, the DBMS applies that model at runtime. Because the model is being applied at runtime, inference time matters, as we do not want to unnecessarily slow down query processing. Next, we discuss the inference time when these models are trained using our learned operator embeddings. We also performed a Pareto analysis of the efficiency (inference time) vs.~the effectiveness (accuracy) of these models.

\sparagraph{Inference time} Figure~\ref{fig:throughput} shows the amount of time it takes to perform model inference on a single operator for each machine learning model we used and feature engineering technique. Note the log scale on the y-axis. The \sparse featurization has exceptionally high inference time, especially for kNN and SVM models, because these models require measuring the distance between the input vector and a large number of other points (for kNN, this could potentially be every point in the training set; for SVM, this could be a large number of support vectors).

The \pca and \fa featurizations have slightly lower inference time than the \neural featurization. This is because running the sparse feature vector through the multi-layer deep neural network takes slightly longer than the simple dot product computations required for \pca and \fa. However, the difference is minimal ($< 1$ms). It is important to note that operator embeddings (and other dimensionality reduction techniques) lead to \emph{faster inference time than the original sparse vector encoding}, because the embedding process is often asymptotically cheaper than the inference process. This is especially true for models such as SVM and kNN, which scale poorly with dimensionality -- for example,  inference time using \sparse featurization with an SVM model (130ms) is almost an order of magnitude more than using the \neural featurization (18ms).

\sparagraph{Pareto analysis} Of course, inference time is not the only important factor when considering a runtime model: accuracy is obviously important as well. Here, operator embeddings' natural synergy with logistic regression (see Section~\ref{sec:admit}) give the \neural featurization a massive advantage, as illustrated in the Pareto plot in Figure~\ref{fig:pareto}.

Figure~\ref{fig:pareto} shows all the models trained for the cardinality boosting task for the TPC-DS dataset, plotted based on their inference time (x-axis, log scale) and their testing accuracy (y-axis). The Pareto front, the models for which no other model is both faster \emph{and} more accurate, contains only the logistic regression classifiers using the \neural, \fa, and \pca featurizations. While the logistic regression classifiers using the \fa and \pca featurizations have slightly faster inference time (the \neural featurization still produces an inference time under 0.1ms), the accuracy of the model produced by the \neural classifier is \emph{substantially} higher: approximately 94\% versus 73\%. Therefore, with the exception of the two logistic regression models with significantly lower accuracy, we conclude that \emph{the models produced by the learned operator embeddings are Pareto dominate in terms of inference time and accuracy.}

\subsection{Training Time}
{Finally, we analyze the time required to train the hourglass-shaped neural network used to represent an embedding. We compare the training time required for both a CPU (Intel Xeon E5-2640 v4) and a GPU (GeForce GTX TITAN). The results are shown in Figure~\ref{fig:runtime}. The time to train the network is a function of several parameters, the most notable being the size of the training set. Since our TPC-DS workload has the most queries (21,688), it has a longer training time than the online (8,000 queries) or batch (1,500 queries) workloads. For TPC-DS, the largest workload, training time on a CPU required 10 minutes, whereas on a GPU the training time dropped to 2.8 minutes.

Since the learned embedding does not need to be frequently retrained (as demonstrated by the consistent model performance on the online workload), we conclude that \emph{the training overhead of the proposed learned embeddings is manageable even for systems without a GPU}, although we note that a GPU greatly accelerates training. Of course, training time can be reduced by sampling from the available training set, and we leave studying the effects of such a strategy on model performance to future work.}


%% file: related_work.tex
\section{Related work}
\label{sec:related}

\sparagraph{Machine learning in DBMSes} A recent groundswell of work has focused on applying machine learning to data management problems. Recent works in intrusion detection~\cite{dbsafe, cyberrank}, index structures~\cite{ml_index}, SLA management~\cite{perfenforce_demo, wisedb-vldb, wisedb-cidr, step, slaorchestrator, mdp_elastic, delphi_pythia}, entity matching~\cite{deep_entity}, physical design~\cite{nashdb, selfdrivingcidr}, and latency prediction~\cite{opt_est_par, LiRobustestimationresource2012, opt_est, pred_multiple, jennie_sigmod11, learning_latency, ernest, contender} have all employed machine learning techniques. SageDB~\cite{sagedb} proposes building an entire database system around machine learning concepts, including integrating machine learning techniques into join processing, sorting, and indexing. With little exception, each of these works have included hand-engineered features derived for each particular task, a arduous process.

Notably, recent work has applied reinforcement learning to various problems in data management systems, including join order enumeration~\cite{rejoin, cidr_dlqo, sanjay_wat}, cardinality estimation~\cite{qo_state_rep}, and adaptive query processing~\cite{adaptive_qp_rl, cuttlefish, skinnerdb}. Recent work~\cite{deep_card_est, deep_card_est2} has also used more traditional supervised learning approaches, using a specialized neural network architecture, to perform join cardinality estimation. Predominately, these techniques are custom-tailored to a problem at hand, and while several of the systems mentioned utilize deep learning (and thus learn useful features automatically for their specific task), they do not generally decrease the burden of applying machine learning to \emph{new} problems in data management.

\sparagraph{Feature engineering} Recent works related to feature engineering, such as ZOMBIE~\cite{zombie}, Brainwash~\cite{brainwash}, and Ringtail~\cite{ringtail} take a ``human-in-the-loop'' approach to feature engineering, assisting data scientists in selecting good features. {These techniques all seek to automate or shorten the process of evaluating the utility of a particular feature (e.g., optimizing model testing), a time-consuming task in feature engineering.}  In contrast, our technique takes a completely automatic approach, custom-tailoring a featurization to a specific database without any user interaction.

\sparagraph{Automatic machine learning} As machine learning grows more popular and complex, recent research has also focused on automating the entire machine learning pipeline~\cite{northstar, helix, BinnigInteractiveCurationAutomatic2018}. These systems are generally designed for use by data science practitioners applying machine learning to external problems, and assist data scientists with model selection, hyperparameter tuning, etc. In contrast, our approach focuses on automatically generating features for machine learning applications \emph{within} the DBMS and with generating features in an \emph{entirely} automated fashion.

\sparagraph{Learned embeddings} {This work is not the first to apply learned embeddings to database systems. In~\cite{deep_entity}, the authors show how embeddings learned on the rows of a table (as opposed to this work, which learns an embedding of query operators) can be used for entity matching. In preliminary work~\cite{termite}, the authors present Termite, a system that helps users navigate and explore heterogeneous databases by building a multi-faceted embedding on structured and unstructured data. In~\cite{sql_embed}, the authors present a method for embedding the text of SQL queries (the actual query as written by the user) using recurrent neural networks, and show that the features learned are useful for various textual tasks (such as detecting syntax errors).}   Most researching into embeddings have focused on natural language processing~\cite{word2vec, SnyderDeepneuralnetworkbased2016} or computer vision~\cite{caffe}. In contrast, our work demonstrates the power of deep learning as a feature engineering tool for operator-level data management tasks.


%% file: conclusion.tex
\section{Conclusions}
\label{sec:conclusions}

We have presented flexible operator embeddings, an automatic technique to custom-tailor a multi-purpose featurization to a particular database. Utilizing deep learning, our technique learns semantically rich, information dense embeddings of query operators automatically, and with minimal human interaction. We have shown that our technique produces features that can be utilized by simple, well-studied, off-the-shelf machine learning models such as logistic regression, and that the resulting trained models are both accurate and fast.

Moving forward, we plan to investigate additional applications of operator embeddings, especially in parallel databases. We are also considering new ways to integrate the embedding training process into modern query optimizers, and if there is any way the training process could exploit recorded partial execution information about past queries.


%% file: apx_encoding.tex
\begin{table}
  \centering
  \scalebox{0.45}{
    \tabcolsep=0.11cm
\begin{tabular}{cccl}
\toprule
  Feature             & PostgreSQL Ops & Encoding & Description \\ \midrule
  Plan Width          & All                & Numeric  & Optimizer's estimate of output row width \\
  Plan Rows           & All                & Numeric  & Optimizer's cardinality estimate \\
  Plan Buffers        & All                & Numeric  & Optimizer's estimate of operator memory requirements \\
  Estimated I/Os      & All                & Numeric  & Optimizer's estimate of the number of I/Os performed \\
  Total Cost          & All                & Numeric  & Optimizer's cost estimate for this operator, plus the subtree \\ \midrule
  Join Type           & Joins              & One-hot  & One of: semi, inner, anti, full \\ 
    Parent Relationship & Joins              & One-hot  & When the child of a join. One of: inner, outer, subquery \\ 
  Hash Buckets        & Hash               & Numeric  & \# hash buckets for hashing \\ 
  Hash Algorithm      & Hash               & One-hot  & Hashing algorithm used \\
    Sort Key          & Sort                & One-hot  & Key for sort operator\\
  Sort Method         & Sort              & One-hat  & Sorting algorithm, e.g. ``quicksort'', ``top-N'', ``external sort''\\ \midrule
  Relation Name       & All Scans          & One-hot  & Base relation of the leaf \\
  Attribute Mins      & All Scans          & Numeric  & Vector of minimum values for relevant attributes \\
  Attribute Medians   & All Scans          & Numeric  & Vector of median values for relevant attributes \\
  Attribute Maxs      & All Scans          & Numeric  & Vector of maximum values for relevant attributes \\
  Index Name          & Index Scans         & One-hot  & Name of index \\
  Scan Direction      & Index Scans        & Boolean  & Direction to read the index (forward or backwards) \\ \midrule
  Strategy            & Aggregate         & One-hot  & One of: plain, sorted, hashed \\
  Partial Mode        & Aggregate          & Boolean  & Eligible to participate in  parallel aggregation \\
  Operator            & Aggregate          & One-hot  & The aggregation to perform, e.g. \texttt{max, min, avg} \\
\end{tabular}
}
\vspace{3mm}
\caption{Features used for naive encoding}
\label{tbl:features}
\end{table}

\section{Encoding from PostgreSQL}
\label{apx:encoding}
Here, we describe the naive, sparse encoding we derive from the PostgreSQL \texttt{EXPLAIN} output.

Table~\ref{tbl:features} describes the values used for our naive encoding. The first column lists the name of the quantity. The second column describes which PostgreSQL operators use a particular type of input. The third column describes how the particular value is encoded into an input suitable for a neural network. The encoding strategies are:
\begin{itemize}
\item{{\bf Numeric}: the value is encoded as a numeric value, occupying a single vector entry.}
\item{{\bf Boolean}: the value is encoded as either a zero or a one, occupying a single vector entry.}
\item{{\bf One-hot}: the value is categorical, and is encoded as a one-hot vector, e.g. a vector with a single ``1'' element where the rest of the elements are ``0'', occupying a number of vector entries.}
\end{itemize}

A particular operator is encoded using all of its applicable values, and using zeros for all inapplicable values. For example, a join operator will be encoded as a vector that has zeros for the ``Strategy,'' ``Partial Mode,'' and ``Operator'' entries, as these apply only to aggregate operators. An aggregate operator, on the other hand, would have zeros for the ``Join Type'' (for example) entry. See Section~\ref{sec:embedding} for additional details on the sparse encoding.
